\documentclass[aps,prb,twocolumn,showpacs,longbibliography,nofootinbib,superscriptaddress,10pt]{revtex4-2}
\usepackage{graphicx,mathtools,bm,array,enumitem,overpic,booktabs,multirow,pifont,etoolbox,amsmath,amssymb,amsfonts,appendix}
\usepackage{natbib}
\usepackage[dvipsnames]{xcolor}
\usepackage{braket}
\usepackage{textcomp}
\usepackage{comment}
\usepackage{chngcntr}
\usepackage[bookmarks=true,colorlinks,linkcolor=RoyalBlue,urlcolor=RoyalBlue,citecolor=RoyalBlue]{hyperref}
\usepackage{cleveref}
\usepackage{orcidlink}
\usepackage{tikz}
\usepackage{soul}

\usepackage{cellspace} 
\begin{document}
\title{Plasmon decay and non-equilibrium steady states in Josephson junction chains}
\author{Lucia Vigliotti${}^{\orcidlink{0000-0002-7496-9717}}$}
\affiliation{Institute of Science and Technology Austria (ISTA), Am Campus 1, 3400 Klosterneuburg, Austria}
\author{Andrew P. Higginbotham${}^{\orcidlink{0000-0003-2607-2363}}$}
\affiliation{James Franck Institute and Department of Physics, University of Chicago, 929 E 57th St, Chicago, Illinois 60637, USA}
\author{Maksym Serbyn${}^{\orcidlink{0000-0002-2399-5827}}$}
\affiliation{Institute of Science and Technology Austria (ISTA), Am Campus 1, 3400 Klosterneuburg, Austria}
\date{\today}

\begin{abstract}
Josephson junction (JJ) chains combine the coherence of superconductivity with the controllability of microwave-frequency circuits, making them a powerful platform for circuit quantum electrodynamics. In this work we consider a long JJ chain that effectively realizes a multi-mode cavity with nonlinear dispersion and additional multi-mode interactions. Individual modes appearing due to the finite size of the chain can be experimentally probed via microwave spectroscopy, both in equilibrium and in driven far-from-equilibrium settings. We study the role of multi-mode interactions in degrading internal coherence -- observable as excess linewidth -- in both equilibrium and driven regimes. Focusing on two-into-two mode scattering as the leading relaxation process, we classify the relevant scattering processes and derive their expected temperature- and frequency-scaling under equilibrium conditions. For experimentally relevant parameters, we show that the equilibrium decay rate is dominated by non-resonant processes, however weakly driving a particular set of modes out of equilibrium enhances resonant scattering, leading to observable signatures in the distribution function and linewidth. Finally, in the strong non-equilibrium regime we report a crossover to a qualitatively different non-equilibrium steady state. 
\end{abstract}

\maketitle

\section{Introduction}
Building on early theoretical work on macroscopic quantum phenomena~\cite{leggett_ptp_1980,leggett_prb_1984} and a series of pioneering experiments in superconducting devices~\cite{turutanov_lowtempphys_2025}, sustained research interest in superconducting circuits emerged in the mid-1980s, when their simplest realization -- based on a single Josephson junction (JJ) --  enabled clear and controllable demonstrations of macroscopic quantum tunneling and energy quantization~\cite{devoret_prl_1985,clarke_science_1988}. Over the years, JJs have been extensively explored as building blocks for qubits, memory, and processors in quantum science~\cite{devoret_arxiv_2004,matanin_prapp_2023,naik_ncomms_2017,koch_pra_2007}. It soon became evident that assembling many junctions into arrays not only provides further potential for applications in quantum technologies~\cite{pop_nature_2014,manucharyan_science_2009,masluk_prl_2012}, but gives access to a rich set of low-energy, near-equilibrium properties, arising from the competition between ground-state configurations and their lowest excitations. Among these, the superconductor-insulator transition~\cite{chow_prl_1998,kuzmin_natphys_2019,mukhopadhyay_natphys_2023}, frustration effects~\cite{vanotterlo_prb_1993,pernack_prb_2024}, vortex dynamics~\cite{fazio_physrep_2001}, quantum phase slips~\cite{rastelli_prb_2013,pop_natphys_2010,ergul_njp_2013,ergul_scirep_2017}, and the synthesis of interesting models such as two-dimensional XY spin models~\cite{vanwees_prb_1987}, spin glasses~\cite{chandra_prl_1995} and spin liquids~\cite{chamon_prl_2020}.

Advances in fabrication techniques allow to realize JJ chains in different regimes, and to probe their behavior with modern spectroscopic methods under both equilibrium and non-equilibrium conditions. In particular, pioneering works considered JJ chains with very weak bulk nonlinearity and a strong nonlinear impurity placed at the boundary~\cite{kuzmin_prl_2021,burshtein_prl_2021,mehta_nature_2023}. This setup was used to study phase slips scattering~\cite{kuzmin_prl_2021,burshtein_prl_2021} and also probe multi-mode interactions induced by the boundary impurity~\cite{mehta_nature_2023}. In contrast, the effect of \emph{bulk nonlinearity}, where multi-mode scattering is allowed throughout the JJ chain, received less attention, especially in out-of-equilibrium settings.  

The situation where a JJ chain is homogeneous and features considerable bulk nonlinearity was recently experimentally studied by the present authors~\cite{bubis_sciadv_2026}. The experiment employs microwave spectroscopy to resolve individual collective modes in the JJ chain, characterize their linewidth and study their out-of-equilibrium behavior.  This raises natural questions: How do intrinsic many-body processes compete with the extrinsic decoherence due to environment coupling, potentially impacting device performance? How can we change the balance between intrinsic and extrinsic decoherence by subjecting the system to non-equilibrium conditions? What universal behavior can emerge far from equilibrium in this system? 

In order to provide theoretical insights into the above questions, in this work we lay down the translation of JJ chain physics in the superconducting regime to the well-established cavity language \cite{aspelmeyer_rmp_2014}. We set up a kinetic equation to calculate the steady state of the system, and study its fate without driving and in presence of a strong driving, that was demonstrated experimentally to stimulate the nonlinear multi-mode scattering \cite{bubis_sciadv_2026}.  Without driving, we go beyond previous works \cite{lin_prl_2013, bard_prb_2018}, which assumed a continuum of modes, by including the discrete nature of modes and their extrinsic broadening. Importantly, the latter enables processes which would otherwise be off-resonant, which we include in our analysis. Indeed, mode decay is governed by the competition between the energy mismatch introduced by the nonlinear dispersion and the finite linewidth associated with a limited quality factor. Whereas previous works considered the regime where the dispersion-induced energy mismatch exceeds the linewidth, experimentally relevant systems lie in the complementary regime, where off-resonant processes dominate the kinetics for realistic system parameters. We group the relevant scattering processes in two classes and in the equilibrium regime we obtain the characteristic scaling of intrinsic decay rate with temperature and mode number, resulting in predictions that can be verified in experiments~\cite{bubis_sciadv_2026}.

\begin{figure*}[tb]
\centering
        \includegraphics[width =\textwidth]{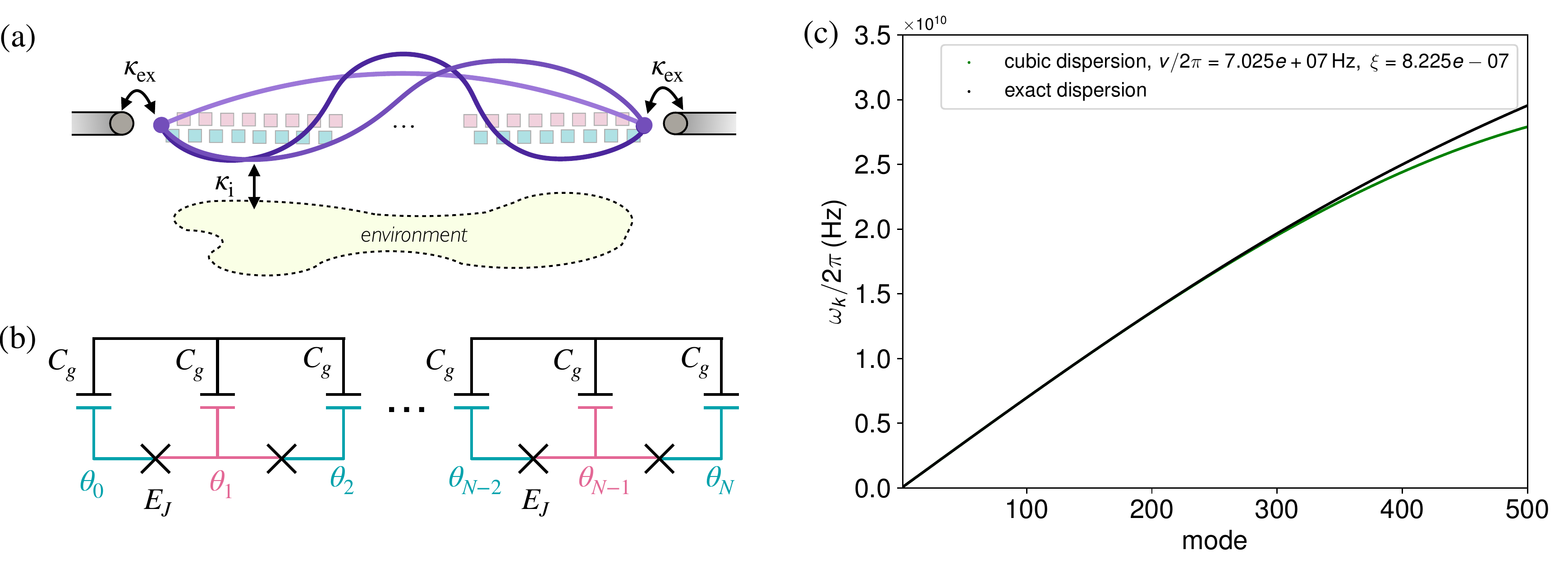}
        \caption{(a) Sketch of system, consisting of a chain of JJs coupled to a transmission line. Small squares denote adjacent superconducting islands. Pink and cyan are used to distinguish adjacent islands, as in (b). Lines in different shades of magenta color illustrate the wave function of the first plasmon modes with number $k=1,2,3$. The coupling to terminals $\kappa_\text{ex}$ and to the internal environment $\kappa_\text{i}$ are also shown. (b) Detail of the circuit, with $E_J$ the Josephson energy, $C_g$ the capacitance to ground, and $\theta_i$ the superconducting phase of each island. (c) Comparing exact, Eq.~\eqref{eqn:exactdisp}, and expanded, Eq.~\eqref{eqn:mirlindisp} dispersion relation of the chain shows agreement for low-lying modes.}\label{fig:1}
\end{figure*}

In the second part of our work, we incorporate the effect of driving some of the JJ chain modes with a microwave field. This brings the cavity into a non-equilibrium steady state (NESS) where drive and losses are balancing each other. We consider different configurations of modes that are driven: for weak to moderate driving of low-energy modes along with a particular higher energy mode, we find an enhancement of resonance processes that were practically unobservable in the equilibrium setting. Next, under moderate driving of only the higher energy mode we find linewidth narrowing of nearby modes caused by the incoming flux of excitations. Finally, in the regime of strong driving applied solely to low energy modes, enhanced intrinsic scattering leads to the emergence of a qualitatively different steady state. In this state the system no longer retains memory of the pumping configuration, and the characteristic dependence of modes' linewidth, dominated by intrinsic interactions, features a qualitatively different scaling on the driving power. 

Our theoretical results provide insights into scalings observed in experiments~\cite{bubis_sciadv_2026}, and also encourage a systematic experimental study of the distribution function and its response to increasing driving power. Beyond its fundamental interest, our analysis yields practical guidance for device engineering: unless the drive is strong, we find that intrinsic thermalization of collective modes in JJ chains is slow and is not the limiting factor. As a consequence, we expect that even longer and lower loss JJ chains will continue to serve as a linear, bosonic bath for quantum simulation~\cite{kuzmin_prl_2021,burshtein_prl_2021,mehta_nature_2023} and Bloch physics~\cite{crescini_natphys_2023,kuzmin_natphys_2025} without being limited by intrinsic decoherence. Moreover, our work underlines the role of nonlinear two-into-two scattering processes and opens the door to their potential use to establish quantum inter-mode coherences~\cite{bourassa_pra_2012,imamoglu_prl_1997,mukhin_supsciandtech_2013}.

The remainder of this paper is organized as follows. In Section~\ref{Sec:methods}, we introduce the system's Hamiltonian, including the nonlinear interaction, the kinetic equation describing its evolution in presence of extrinsic damping, multi-mode scattering and a drive, and we establish a connection between the JJ chain parameters and the cavity language. In Sections~\ref{sec:thermal}-\ref{sec:noneq} we present our results, starting with analytic calculations of the equilibrium decay rate in Sec.~\ref{sec:thermal}, and extending them to the non-equilibrium regime in Sec.~\ref{sec:noneq}. Finally, in Section~\ref{sec:disc} we summarize our results and discuss open questions and potential avenues for future work.

\section{System and methods \label{Sec:methods}}
In this Section, we introduce the theoretical description of the JJ chain (Sec.~\ref{subsec:methodsA}) and its evolution through kinetic equation (Secs.~\ref{subsec:kinthermal}). Finally, in Sec.~\ref{subsec:parameters} we discuss the mapping between the JJ chain and the multi-mode cavity language and accessible values of experimental parameters. 

\subsection{From the non-interacting Hamiltonian to four-wave mixing}\label{subsec:methodsA}
We study a one-dimensional chain of $N$ JJs for $N+1$ superconducting islands, labeled by $n,\,m=0,\dots,N$, coupled on either side to a microwave transmission line, see Fig. \ref{fig:1}(a). Neighboring superconducting islands are coupled via Coulomb interaction between their charges $Q_n$ and via Josephson effect between their phases $\theta_n$~\cite{krupko_prb_2018,basko_prb_2020,muppalla_phdthesis_2020}:
\begin{multline}
    H=\frac{1}{2}\sum_{n,m=1}^{N-1}Q_nC^{-1}_{nm}Q_m-E_J\sum_{n=1}^{N-2}\cos(\theta_{n+1}-\theta_n)\\-E_J\cos \theta_1-E_J\cos \theta_{N-1} .\label{eqn:krupko}
\end{multline}
The first term is the charging energy and the second one is the nonlinear Josephson potential, with $E_J=(\hbar/2e)^2/L_J$, $L_J$ being the inductance of a single junction. The two variables $Q_n$ and $\theta_n$ are canonically conjugated, $[\theta_n,Q_m]=\mathrm{i}2e\delta_{n,m}$. Due to the low impedance of the external circuit, the phase is effectively pinned at the boundaries, and we impose $\theta_0=\theta_N=0$. Notice that such constraint is responsible for the limits 1 and $N-2$ of the second sum in Eq.~(\ref{eqn:krupko}). Regarding the charging part, $Q_n$ are the canonical variables conjugate to the phases, but it is sufficient to define them only for the phases that are actual dynamical degrees of freedom, namely $\theta_1, \ldots, \theta_{N-1}$, given the others are set to zero. For this reason, the charging sum runs from $1$ to $N-1$. In the simplest case, each junction is only coupled to ground with on-site capacitance $C_g$ and to its nearest neighbors with capacitance $C_J$, such that the capacitance matrix $C_{nm}$ is tridiagonal,
\begin{align}
C_{nm}
=C_g\delta_{n,m}+C_J(2\delta_{n,m}-\delta_{n,m+1}-\delta_{n,m-1}).
\end{align}
In Fig. \ref{fig:1}(b) we show details of the circuit.

If the Josephson energy is much larger than the charging energy, the phase is only weakly fluctuating around the superconducting ground state (where all $\theta_n$ would be a constant, equal to 0 for our boundary conditions), and the nonlinear Josephson potential can be expanded around $\theta_{n+1}-\theta_n=0$. This expansion amounts to ignoring phase slips~\cite{halperin_intjmodphysb_2010,haviland_natphys_2010}; we return to this approximation at the end of this Section and also in Section~\ref{sec:disc}. Moreover, since the charging energy is much smaller than the Josephson energy, the system is largely insensitive to background charges, which justifies neglecting offset charges in Eq.~\eqref{eqn:krupko}. At the lowest (second) order, we get
\begin{multline}
    H^{(2)}=\frac{1}{2}\sum_{n,m=1}^{N-1}Q_nC^{-1}_{nm}Q_m+\frac{E_J}{2}\sum_{n=1}^{N-2}(\theta_{n+1}-\theta_n)^2\\+\frac{E_J}{2}\theta_1^2+\frac{E_J}{2}\theta_{N-1}^2.\label{eqn:krupkolin}
\end{multline}
From the circuit point of view, this corresponds to considering each junction as a linear component, made by a capacitance $C_J$ in parallel with an inductance $L_J$.

We first diagonalize $H^{(2)}$ in mode space, and then second quantize it. We introduce the discrete quasi-momenta $\pi k/N$, where $k\in\mathbb{N}$ is the mode number labeling the excitations. We express distances in units of the junctions' spacing $L/N$, where $L$ is the chain's size, whence quasi-momenta are dimensionless. We start by expanding both $\theta_n$ and $Q_n$ in normal modes,
\begin{equation}
    \theta_n=\sum_{k=1}^{N-1}\psi_k(n)\theta_k,\qquad Q_n=\sum_{k=1}^{N-1}\psi_k(n)Q_k,
\end{equation}
where
\begin{equation*}
    \psi_k(n)=\sqrt{\frac{2}{N}}\sin{\left(\frac{\pi k n}{N}\right)},\quad k=1,\dots,N-1
\end{equation*}
encode the distribution of charge and phase along the chain for a given mode $k$ (wavefunctions in a continuum model). Generally, the $k=0$ mode corresponds to a uniform shift of all island phases, i.e. the global superconducting phase. Since the Josephson energy depends only on phase differences, this is a zero-eigenvalue mode of the Josephson energy. Its dynamics is governed solely by the charging energy and does not represent a propagating collective excitation. However, the boundary conditions $\theta_0=\theta_N=0$ fix the global phase and therefore eliminate this mode from the spectrum. All $\psi_k(n)$ satisfy open boundary conditions, $\psi_k(0)=\psi_k(N)=0$, and
are orthogonal to each other, $\sum_{n=1}^{N-1}\psi_k(n)\psi_{k'}(n)=\delta_{k,k'}$. After substitution in Eq.~\eqref{eqn:krupkolin}, and introducing $\phi_k=\hbar \theta_k/(2e)$ which are the momentum-space equivalent of the node flux for each island, we get
\begin{equation}
    H^{(2)}=\sum_{k=1}^{N-1}\left[\frac{1}{2C_k}Q_k^2+\frac{L_k^{-1}}{2}\phi_k^2\right],\label{eqn:harmosc}
\end{equation}
with mode-dependent effective parameters ${C_k=C_g+2C_J\left[1-\cos{\left(\pi k/N\right)}\right]}$ and ${L_k^{-1}=2E_J(2e)^2/\hbar^2\left[1-\cos{\left(\pi k/N\right)}\right]}$, which have units of capacitance and inductance.

Given the commutation relation $[\phi_k,Q_{k'}]=\mathrm{i}\hbar\delta_{k,k'}$, we recognize Eq.~\eqref{eqn:harmosc} as a set of decoupled harmonic oscillators where $\phi_k$ plays the role of position and $Q_k$ the role of momentum. The corresponding frequencies are given by \cite{masluk_prl_2012}
\begin{equation}
    \omega_k=\sqrt{\frac{1}{L_kC_k}}=\omega_P\sqrt{\frac{1-\cos{\left(\frac{\pi k}{N}\right)}}{\frac{C_g}{2C_J}+1-\cos{\left(\frac{\pi k}{N}\right)}}},\label{eqn:exactdisp}
\end{equation}
where $\omega_P=1/\sqrt{L_JC_J}=\sqrt{2E_JE_c}/\hbar$ has units $\text{rad}\cdot \text{s}^{-1}$ and is the plasma frequency, with $E_c=(2e)^2/(2C_J)$. Intuitively,  each mode $k$ describes a long-wavelength fluctuation of the phase degree of freedom along the JJ chain.

We can express Eq.~\eqref{eqn:harmosc} in a second quantized form. Following the analogy with the harmonic oscillator, we introduce raising and lowering operators
\begin{equation}
    \theta_k=2e\sqrt{\frac{1}{2\hbar C_k\omega_k}}\left(\hat{a}_k+\hat{a}_k^{\dagger}\right).
\end{equation}
The second quantized Hamiltonian reads
\begin{equation}
H^{(2)}=\sum_{k=1}^{N-1}\hbar\omega_{k}\left(\hat{a}_{{k}}^{\dagger}\hat{a}_{{k}}+\frac{1}{2}\right),
\end{equation}
where $\hat{a}^{\dagger}_k\,(\hat{a}_k)$ are bosonic operators that create (annihilate) a plasmon excitation in the $k$-th mode.

In the following, we will use a simplified expression for the phase field $\theta_n$~\cite{bard_prb_2019}, obtained in the limit $k\ll N$. At the lowest order in $k/N$, the frequency spectrum is linear and the capacitance parameter is simplified,
\begin{equation}
    \omega_k \approx v k, \quad C_k\approx C_g,
\end{equation}
with $v=\sqrt{2E_JE_g}\pi/(\hbar N)$ and $E_g=(2e)^2/(2C_g)$. Then $\theta_n$ becomes\footnote{In this limit, the continuum version of the Hamiltonian can be mapped onto a Luttinger liquid, with $K_\mathrm{g}$ defined below being the corresponding Luttinger constant, see Refs.~\cite{bard_prb_2018,houzet_prl_2019}.}
\begin{equation}
    \theta_n\approx\frac{1}{\sqrt{\pi K_g}}\sum_{k=1}^{N-1} \frac{1}{\sqrt{k}}\sin{\left(\frac{\pi k n}{N}\right)} (\hat{a}_k + \hat{a}_k^{\dagger}),\label{eqn:thetaLutt}
\end{equation}
with $K_\mathrm{g}=\sqrt{{E_J}/({2E_g})}$ the global superfluid phase stiffness.

We are interested in multi-mode scattering and relaxation arising from the fourth order expansion terms of the full Hamiltonian, denoted as $H^{(4)}$. A perturbative treatment of these terms within the linear approximation is not possible, as the decay rate of plasmons would diverge, see~\cite{imambekov_rmp_2012} and references therein. It is therefore essential to include the finite curvature of the spectrum, which regularizes the divergence. We will do it at the level of transition probabilities (see Sec.~\ref{subsec:kinthermal}). The dispersion relation with the nonlinear correction reads
\begin{equation}
    \omega_k= k(v-v\xi k^2),\label{eqn:mirlindisp}
\end{equation}
where $\xi=\pi^2 E_g/(2E_c N^2)$. With a choice of parameters compatible with current realizations of JJ chains (see Table \ref{table:parameters}), the exact dispersion Eq.~\eqref{eqn:exactdisp} and its cubic expansion Eq.~\eqref{eqn:mirlindisp} are shown in Fig.~\ref{fig:1}(c). Up to $k_\text{max}=500$, they differ by less than $6\%$. Although this discrepancy may exceed the linewidth for high mode numbers, therefore altering their resonant conditions (see Sec.~\ref{subsec:kinthermal}), it conveniently allows analytical tractability of the system in the first part of this work. We have checked that using Eq.~\eqref{eqn:mirlindisp} does not substantially affect our results obtained in equilibrium. For the out-of-equilibrium simulations, we use full Eq.~\eqref{eqn:exactdisp}.

\begin{table}[b]
\begin{ruledtabular}
	\begin{tabular}{ccccc} 
	$E_J/h$ & $E_c/h$ & $E_g/h$ & $N$ & $\kappa^0/2\pi$ \\
	50~GHz & 30~GHz & 0.5~THz & 10000 & 5~MHz\\ 
    \end{tabular}
\end{ruledtabular}
    \caption{Chain parameters.}
	\label{table:parameters}
\end{table}

To switch on mode-to-mode interactions, we reintroduce the quartic order of the nonlinear term, that we will treat perturbatively,
\begin{equation}
  H^{(4)}=-\frac{E_J}{4!}\sum_{n=1}^{N-2}(\theta_{n+1}-\theta_n)^4.\label{eqn:quartic}
\end{equation}

Using Eq.~\eqref{eqn:thetaLutt}, this becomes (see Appendix~\ref{appendixA} for details)
\begin{multline}
        H^{(4)}=- \frac{E_g\pi^2}{96N^3}\sum_{k,p,q_1,q_2>0}\sqrt{kpq_1q_2}(\hat{a}_{k}^{\dagger} + \hat{a}_{k})(\hat{a}_{p}^{\dagger} + \hat{a}_{p})\\(\hat{a}_{q_1}^{\dagger} + \hat{a}_{q_1})(\hat{a}_{q_2}^{\dagger} + \hat{a}_{q_2})\sum_{s_1,s_2,s_3=\pm}\delta_{k+s_1p+s_2q_1+s_3q_2,0}.\label{eqn:nl_2ndq}
\end{multline}
The Kronecker delta, which implements the constraint of quasi-momentum conserving interaction, naturally arises from the wavefunctions overlap. Although mode numbers are strictly speaking positive integers, every sign combination is allowed because each plasma wave supports two components propagating in opposite directions, $e^{\pm ik\pi n/N}$.

Retaining only particle number-conserving terms, the second quantized Hamiltonian can be written in the form (see Appendix~\ref{appendixB} for details on the calculation)
\begin{multline}
H^{(2)}+H^{(4)}=
 \sum_{k} \hbar\omega'_k \hat{a}_k^\dagger \hat{a}_k\\-\frac{\hbar}{2}\sum_k\mathcal{K}_{k,k}\hat{a}_k^\dagger \hat{a}_k\hat{a}_k^\dagger \hat{a}_k
-\frac{\hbar}{2}\sum_{k,p\neq k} \mathcal{K}_{k,p} \hat{a}_{k}^\dagger \hat{a}_k \hat{a}_p^\dagger \hat{a}_p\\
+
\sum_{\substack{
k,p\\
q_1,q_2\neq\{k,p\}
}} 
\mathcal{K}_{k,p,q_1,q_2} \hat{a}_{k}^\dagger \hat{a}^\dagger_p \hat{a}_{q_1} \hat{a}_{q_2},
\end{multline}
with $\omega_k'=\omega_k-\sum_p\mathcal{K}_{k,p}/2$. Here the second line represents self- and cross-Kerr shifts, and the third line a four-wave mixing process, and we excluded self-decay processes that do not represent a two-into-two scattering. Kerr-shifts describe the frequency shift of mode $k$ that scales linearly either with the occupation number in the same mode ($\propto  \hat{a}_{k}^\dagger \hat{a}_k$), or in the other modes ($\propto  \hat{a}_{p}^\dagger \hat{a}_p$). Primed frequencies $\omega'_k$ take into account the renormalization of frequencies due to nonlinearities in the zero occupation number limit (vacuum fluctuations). The Kerr-coefficients read (see Appendix~\ref{appendixB} and Refs.~\cite{weissl_prb_2015,krupko_erratum_2023})
\begin{equation}
    \mathcal{K}_{k,p}=\left(\frac{1}{2}-\frac{\delta_{k,p}}{8}\right)\frac{\hbar \omega_k\omega_p}{2NE_J}.\label{eqn:kerrcoeff}
\end{equation}

In the remainder of the paper, we will neglect Kerr corrections and rather focus on four-wave mixing. We restrict our analysis to number-conserving terms, as previous works have shown them to play a dominant role in the relaxation of bosons in similar contexts~\cite{apostolov_prb_2013,protopopov_prb_2014}. Moreover, two-into-two processes admit on-shell configurations, i.e., channels where momentum and energy resonant conditions are satisfied even in the absence of modes' broadening. Slight deviations from these configurations lead to only weak energy mismatch, compatible with the finite linewidth of the modes. In contrast, one-into-three scattering processes do not have on-shell channels, resulting in unfavorable, strongly off-resonant conditions. More details can be found in Appendix~\ref{appendixC}.

The matrix element of two-into-two scattering reads
\begin{multline}
    \mathcal{K}_{k,p,q_1,q_2}=\langle 0| \hat{a}_k \hat{a}_pH^{(4)} \hat{a}_{q_1}^\dagger \hat{a}_{q_2}^\dagger |0\rangle\\=-\frac{E_g\pi^2}{4N^3} \sqrt{kpq_1q_2}\sum_{s_1,s_2,s_3=\pm}\delta_{k+s_1p+s_2q_1+s_3q_2,0}.\label{eqn:boxmatrixel}
\end{multline}
We checked explicitly that the contribution of higher orders is strongly suppressed (see Appendix~\ref{appendixD}). Since plasmons are probed via excitation with microwave radiation, in the following we will also talk about \textit{photon} scattering. An example of quasi-momentum conserving scattering is shown in Fig.~\ref{fig:notation}(a). The energy conservation constraint will be extensively discussed below. 

To conclude this part, we emphasize that phase slips, mentioned in the introduction, cannot be ruled out \textit{a priori}. These are non-perturbative processes where the phase difference across a junction changes by $2\pi$, and they also contribute to the intrinsic relaxation of modes, which is the focus of our work. However, a recent experiment~\cite{bubis_sciadv_2026}, performed at energy scales comparable to those listed in Table~\ref{table:enscales}, showed that the scaling of intrinsic decay rates was compatible with gradient nonlinearities described by Eq.~\eqref{eqn:quartic}. Previously, Ref.~\cite{bard_prb_2018} predicted that damping due to scattering off quantum phase slips becomes relevant only at frequencies exponentially small in the ratio $E_J/E_c$. In light of these results, we therefore consider gradient nonlinearities as the sole source of intrinsic plasmon decay in the following.

We move now to plasmon relaxation and decay rates. For notational convenience, we set $\hbar=k_B=1$ from now on.

\subsection{Kinetic equation and decay rates}
\label{subsec:kinthermal}
To describe the scattering processes from quartic nonlinearity, we use the framework of a kinetic equation \cite{bard_prb_2019_b,bhattacharyya_prl_2020,fazio_prl_1998}.  The kinetic equation provides a simplified description of the interacting modes, that neglects coherences between different modes (justified, in particular, since we are interested in steady state properties and coupling to environment destroys phase coherence) and assumes that the state of the system is specified by the occupation numbers of individual modes, 
\begin{equation}\label{Eq:nk}
   n_k=\langle\hat{a}^\dagger_k\hat{a}_k\rangle.
\end{equation}
The distribution number evolves in time according to Boltzmann kinetic equation
\begin{equation}
    \frac{\partial n_k}{\partial t}=I_k[n] - \kappa^0 (n_k-n_k^\text{th}) + \kappa_{\text{ex}} n^\text{flux}_k,
    \label{eqn:boltzkineq}
\end{equation}
where the first term, $I_k[n]$, is the collision integral~\cite{apostolov_prb_2013} describing intrinsic scattering processes. The second term describes the coupling to terminals and environment with the effective rate 
\begin{equation}\label{Eq:kappa0}
\kappa^0=2\kappa_{\text{ex}}+\kappa_{\text{i}},
\end{equation} 
see Fig.~\ref{fig:1}(a) and Table~\ref{table:parameters}, that aims to bring modes to thermal equilibrium with Bose-Einstein distribution function $n_k^\text{th}=(e^{\omega_k/T}-1)^{-1}$. Finally, the last term phenomenologically describes the presence of an external driving, with $n^\text{flux}_k$ a power spectral density in units of photons/s/Hz. 

The collision integral depends on the occupation of all modes present in the system, and can be split into an in-scattering part that adds excitations to a certain mode $k$, and an out-scattering part removing excitations, 
\begin{align}
 I_k[n] & =I_k^\text{in}[n]+I_k^\text{out}[n],\\
I_k^\text{in}[n] 
    &=
    \frac{1}{2}\sum_{p,q_1,q_2} W_{q_1, q_2\to p,k} (1+n_p)(1+n_k)n_{q_1}n_{q_2},\label{eqn:collisionin}\\
    I_k^\text{out}[n] 
    &=
    -\frac{1}{2}\sum_{p,q_1,q_2} W_{p,k\to q_1, q_2} (1+n_{q_1})(1+n_{q_2})n_kn_p,
    \label{eqn:collisinout}
    \end{align}
where the additional factor $1/2$ takes into account the permutation symmetry of modes $q_1,q_2$, and we implicitly   exclude unphysical self-decays from the sum ($q_1=k$ or $q_2=k$). 

The transition probabilities $W_{q_1, q_2\leftrightarrow p,k}$ entering into the collision integral are given by the Fermi golden rule, that includes a delta-function to ensure energy conservation. In order to take into account the broadening of modes due to their coupling to environment and terminals, we use the golden rule with a Lorentzian
\begin{equation}
    W_{q_1, q_2\leftrightarrow p,k}
    =2\pi|\mathcal{K}_{k,p,q_1,q_2}|^2\delta_\gamma(\omega_{q_1}+\omega_{q_2}-\omega_{k}-\omega_{p}).
    \label{eqn:Wrate}
\end{equation}
The effective linewidth of a two-into-two scattering process is given by the sum of the original linewidths $\gamma=2\kappa^0$,
\begin{equation}
    \delta_{\gamma}(\Delta\omega)=\frac{1}{\pi}\frac{\gamma}{\gamma^2+(\Delta\omega)^2}\label{eqn:dos}.
\end{equation}
Note that for simplicity we assume that the couplings to the terminals and to the external environment are mode-independent. This is a reasonable assumption, as it is expected for experimental boundary conditions such as in~\cite{bubis_sciadv_2026}. It is straightforward to extend our results to the case where all parameters in Eq.~(\ref{Eq:kappa0}) depend on mode number $k$. 

\begin{figure}[t] 
        \includegraphics[width=\columnwidth]{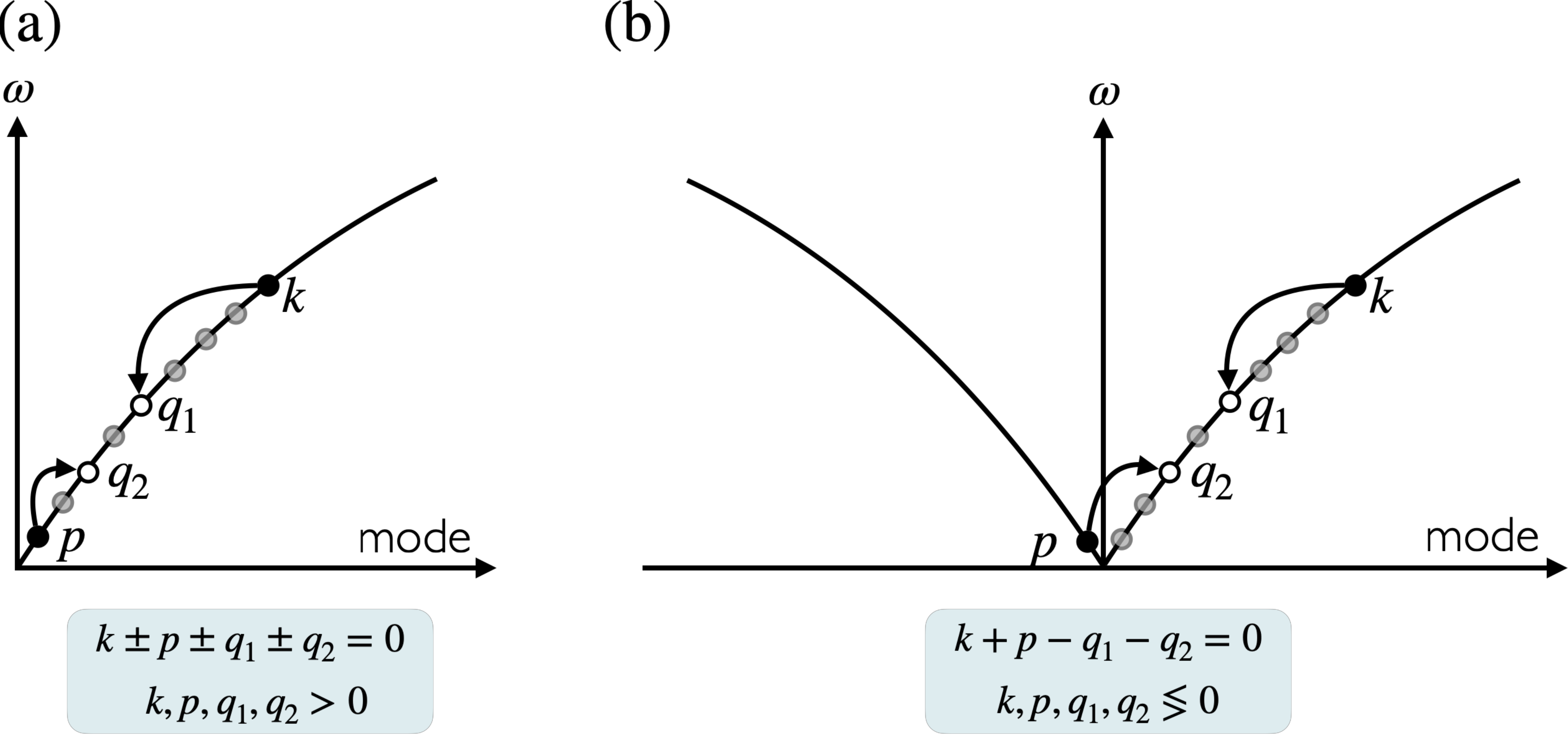}
        \caption{Introducing momentum unfolding: (a) plasmons modes labeled by a set of positive integers, with two-into-two collisions fulfilling the generalized momentum conservation $k\pm p\pm q_1\pm q_2=0$; (b) each mode is split into its positive and negative-momentum component, by introducing negative mode numbers with same frequency and occupation numbers as the positive ones. In this case, the constraint becomes $k+ p-q_1-q_2=0$. In what follows we adopt the unfolded notation as in (b).}\label{fig:notation}
\end{figure}

As we said, the matrix element $\mathcal{K}_{k,p,q_1,q_2}$ in the Fermi golden rule rate, Eq.~(\ref{eqn:boxmatrixel}), also imposes the conservation of momenta. However, due to the chosen open boundary conditions, such momentum conservation is only valid modulo absolute sign, see sketch in Fig.~\ref{fig:notation}(a), as each mode contains two counter-propagating waves with positive and negative momentum.  This more complex momentum conservation may be simplified if we unravel back the dispersion relation, introducing unphysical modes with $k<0$ and imposing a symmetric occupation function, $n_{-k} = n_k$, see Fig.~\ref{fig:notation}(b).  This is a purely theoretical trick, however it simplifies bookkeeping and in what follows we will consider sums over positive and negative momenta, thereby removing the sum over $s_{1,2,3}$ in Eq.~(\ref{eqn:boxmatrixel}).

Within the kinetic equation framework, the effective linewidth of a given mode is defined by linearizing the kinetic equation. In particular, the \emph{intrinsic contribution} to the linewidth stemming from two-into-two scattering is contained in the linearization of the collision integral. The linearization expands the occupation numbers as $n_k=n^0_k+\delta n_k$, where the $n^0_k$ is the NESS of the kinetic equation, satisfying $\partial{n}^0_k/\partial t=0$. In the absence of any external drive, $n^0_k$ are simply given by the Bose-Einstein distribution: the system is at thermal equilibrium with terminals and environment, and Eq.~\eqref{eqn:boltzkineq} describes in practice the microscopic relaxation of mode $k$ in response to a small perturbation $\delta n_k$. Within the relaxation-time approximation, we keep only the diagonal part $\propto \delta n_k$ of the collision integral, ${\partial n_k}/{\partial t}\approx -(\delta\kappa_k +\kappa_0) \delta n_k$,
and obtain an explicit expression for the additional decay rate of mode $k$ due to intrinsic scattering:
\begin{equation}
\delta\kappa_k = \frac{1}{2}\sum_{p,q_1,q_2} W_{p,k\to q_1 q_2}[n^0_p(1+n^0_{q_1}+n^0_{q_2})- n^0_{q_1}n^0_{q_2}],\label{eqn:currentdecays}
\end{equation}
where the sum is subject to the same restrictions as discussed after Eq.~(\ref{eqn:collisinout}). 
The expression~(\ref{eqn:currentdecays}) for intrinsic contribution to linewidth ignores the off-diagonal components of the linearized collision integral. We checked that their contributions are negligible both in equilibrium and in presence of drive, even at high intensity.

The equation~(\ref{eqn:currentdecays}) was used in the previous literature~\cite{lin_prl_2013, bard_prb_2018,bubis_sciadv_2026} to obtain the contribution to equilibrium linewidth from intrinsic scattering. In the next Section we reproduce earlier results and extend them to the regime of broadened modes. In addition, we will use the same expression to get the linewidth in presence of an external driving, $n_k^\text{flux}\neq 0$. In this case, we cannot use the Bose-Einstein distribution function for $n^0_k$. Instead, we numerically determine the NESS of the kinetic equation,  $\partial n^0_k/\partial t=0$, and then use Eq.~\eqref{eqn:currentdecays} to obtain the excess linewidth due to multi-mode interaction.

We notice that the collision integral for many interacting modes features particle and energy conservation. The thermal relaxation and external supply determine the global gain and loss of particles and energy; the collision integral solely redistributes them among modes. Therefore in the driven case we obtain the following expressions for the excitation number and total energy  present in the system, 
\begin{align}
    N_\text{tot} &=  \sum_{k=1}^{N-1} \left[\frac{\kappa_\text{ex}}{\kappa^0}n^\text{flux}_k+n^\text{th}_k\right],\notag\\
E_\text{tot} &=  \sum_{k=1}^{N-1} \omega_k \left[\frac{\kappa_\text{ex}}{\kappa^0}n^\text{flux}_k+n^\text{th}_k\right],\label{eqn:conservation_laws}
\end{align}
which may be used to check for convergence of the kinetic equations and also to define an effective temperature in the system. The NESS results from the balance between the excitation pumping, the losses, and the redistribution of excitations due to the intrinsic collision dynamics.

\subsection{Relevant parameters  and cavity language}\label{subsec:parameters}

Finally, we translate the physical parameters of the JJ chain ($E_g,E_J,E_c,N$ and $\kappa^0$, together with $T$) into the language of multi-mode cavities with nonlinear dispersion, level broadening and two-into-two mode scattering originated by interactions. For convenience, we choose the level spacing of the lowest mode as an overall energy scale, $\Delta_1 =\omega_1 = v$, where $v$ as defined above Eq.~\eqref{eqn:thetaLutt} has energy units, with all the remaining dimensionless parameters discussed below.

\begin{table}[t]
\begin{ruledtabular}
	\begin{tabular}{Sc Sc Sc } 
	Parameter & Scaling & Typical value \\
    \hline
	$\mathcal{F}$ & $\frac{\sqrt{E_gE_J}}{N\kappa^0}$ & $\approx 14$ \\
    $\zeta$ & $\frac{kE_g}{N^2E_c}$ & $\sim 10^{-5}-10^{-3}$\\
    $\zeta \mathcal{F}$ & $\frac{k\sqrt{E_g^3E_J}}{N^3E_c\kappa^0}$ & $\sim 10^{-4}-10^{-2}$\\
    $\bar{p}$ (w/o drive) & $\frac{TN}{\sqrt{E_gE_J}}$ & $\approx 30$\\
	$g$ & $\frac{1}{N^2}\sqrt{\frac{E_g}{E_J}}$ & $\approx 1.8\cdot10^{-8}$\\
    $\tau^\text{min}$ (w/o drive) & $ \frac{E_g^2}{N^6(\kappa^0)^2}$ &$\approx 6\cdot 10^{-14}$\\
	\end{tabular}
    \caption{Parameters that determine the properties of a JJ chain viewed as a multi-mode cavity and their dependence on characteristics of individual JJs, such as their charging ($E_g$, $E_c$) and Josephson ($E_J$) energies, as well as number of junctions $N$, total coupling to the environment $\kappa^0$ and mode number, $k$. When appearing in the above estimates, we use temperature $T=0.1\,\text{K}$ and mode numbers in the range $k=2-200$.}
	\label{table:enscales}
\end{ruledtabular}
\end{table}

The natural ratio of mode width to mode spacing has the intuitive meaning of inverse finesse for the cavity,
\begin{equation}
    \mathcal{F}^{-1}  = \frac{\kappa^0}{\Delta_1} \ll 1,
\end{equation}
that is much smaller than one since typical experiments have well-resolved individual modes~\cite{bubis_sciadv_2026,krupko_prb_2018,weissl_prb_2015}. Next, a convenient dimensionless measure of the cavity dispersion is the ratio
\begin{equation}
    \zeta = \frac{2\omega_k-\omega_{k+1}-\omega_{k-1}}{\Delta_1}.
\end{equation}
This parameter specifies how the level spacing of two-boson decay compares to the single-particle level spacing, and changes along the dispersion. A weak nonlinearity implies that $\zeta \ll 1$, however what matters more is the product $\zeta \mathcal{F}$: 
\begin{equation}
    \zeta \mathcal{F}  = \frac{2\omega_k-\omega_{k+1}-\omega_{k-1}}{\kappa^0}.
\end{equation}
If $\zeta \mathcal{F} \gg 1$, the level broadening is insufficient to overcome the off-resonant mismatch, and two-into-two decay processes remain suppressed. In contrast, once $\zeta \mathcal{F} \lesssim 1$, the finite modes' linewidth becomes large enough to enable these otherwise off-resonant decays. In Table \ref{table:enscales} we will list all the relevant parameters, provide their scaling in terms of the initial parameters of the JJ chain, and estimate their order of magnitude. We see that natural parameters of a JJ array put it in a regime where modes are well resolved, yet the linewidth dominates over non-linearity, enabling incoherent two-into-two mode scattering processes. 

We emphasize that state-of-the-art experiments allow for smaller $\kappa^0$ values than that used in this work, resulting in higher $\mathcal{F}$ and $\zeta\mathcal{F}$. This parameter includes both coupling to the terminals and to the environment (see Eq.~\eqref{Eq:kappa0}). As will be shown, our choice is motivated by the need to discuss all types of off-resonant processes. Smaller values would still allow the off-resonant processes we identify as dominant, and therefore would not alter our conclusions. We also note, that finesse $\cal F$ may be affected by intrinsic decay processes in the cavity, if driven strongly out of equilibrium.

The parameters discussed before can be thought as intrinsic properties of JJ chains, as they become strongly modified only when the system is driven far away from equilibrium state. Now, we discuss two parameters that are strongly influenced by the presence of driving. The first parameter is the (effective) temperature in units of $\Delta_1$, that determines how many modes are occupied by at least one photon,
\begin{equation}
    \bar{p} = \frac{T}{\Delta_1}.
\end{equation}
In the NESS, one can define an effective temperature that would result from matching the total occupation of all modes to the one in a thermal state. 

The second parameter is the thermalization degree $\tau = {\delta \kappa}/{\kappa^0}$. It may be thought as an analogue of the so-called cooperativity \cite{aspelmeyer_rmp_2014}, characterizing how the intrinsic scattering rate $\delta \kappa$ compares to the extrinsic decay rate. The thermalization rate is strongly dependent on details of the distribution function, and the relation between other parameters in the system. However, we can roughly estimate it using the dimensionless matrix element that controls two-into-two scattering process,
\begin{equation} \label{Eq:g-def}
    g = \frac{|\mathcal{K}_{k,p,q_1,q_2}|}{\Delta_1},
\end{equation}
where we used Eq.~(\ref{eqn:boxmatrixel}) and assumed mode numbers to be of order one. Using this notation, the intrinsic scattering rate can be written as $\delta\kappa \propto g^2\Delta_1^2/\kappa^0$, where we replaced DOS by $1/\kappa^0$ (assuming $\zeta \mathcal{F}\leq 1$) and ignored the distribution function and the sum over mode numbers. This leads to the estimate of thermalization rate
\begin{equation}
   \tau = \frac{\delta \kappa}{\kappa^0} \gtrsim \frac{g^2\Delta_1^2}{(\kappa^0)^2} = g^2 \mathcal{F}^2,\label{eqn:therm_par}
\end{equation}
justifying the analogy between $\tau$ and the cooperativity. We emphasize that $\tau$ severely depends on mode numbers involved in the scattering process, and hence the estimate above may be thought as a tentative lower bound. 

When $\tau \ll 1$, the extrinsic decay rate $\kappa^0$ dominates over scattering-induced broadening $\delta\kappa$. In this regime excitations leave the system before they have the opportunity to scatter within the chain, and intrinsic scattering due to nonlinearity is strongly suppressed. In reverse, having $\tau \gtrsim 1$ implies that intrinsic scattering starts to play an important role. We note that, while parameters defined earlier may be thought as approximately constant, $\bar{p}$ and $\tau$ introduced here strongly depend on the distribution function and will be affected by  external pumping. In the regime of strong pumping, the large occupation numbers will enhance intrinsic scattering, leading to a drastic increase of~$\tau$. From Table~\ref{table:enscales}, we see that without driving the thermalization degree $\tau$ is small. However, the parameter $\tau$ effectively serves as a lower bound. After incorporating all relevant factors, it becomes larger, and may be substantially amplified by the drive up to $\tau\sim 10^2$. 

Another useful parameter comparison is between the magnitude of Kerr shifts, and the detuning between modes due to curvature, $\zeta\Delta_1$. Provided that Kerr shifts are larger than $\zeta\Delta_1$, they are capable of enabling or disabling certain intrinsic decay channels. Since, in JJ chains, Kerr shifts and two-into-two particle scattering come from the same mechanism, the former are directly related to the parameters already introduced. Focusing on cross-Kerr (see App. \ref{appendixB} for more details), we get
$\mathcal{K}^{k\neq p}_{k,p}=2g\Delta_1 kp$. Using Table \ref{table:enscales} we see that  $\mathcal{K}^{k\neq p}_{k,p}\ll\zeta\Delta_1$ unless very high frequency modes $p$ are pumped, therefore Kerr shifts will be neglected in our analysis.

The parameters introduced above also allow us to place restrictions on the applicability of the Fermi golden rule, that we use extensively below. The requirement of well-resolved individual modes implies $\mathcal{F}\gg1$, while ensuring that modes do not mix too strongly further requires that $\tau /\mathcal{F} \lesssim 1$. In addition, we can identify  different regimes of Fermi golden rule. When many processes contribute to broadening, the condition $\zeta \mathcal{F}\lesssim 1$ ensures that the linewidth is well averaged over many channels and varies smoothly with mode index. In the opposite mesoscopic regime, $\zeta \mathcal{F}> 1$, some modes have resonant (on-shell) processes while others do not. This leads to a non-smooth dependence of the decay rate on mode number, although the Fermi golden rule remains applicable due to the presence of a bare linewidth.  Finally, the condition $\tau \lesssim 1$ is needed to ensure that no inter-mode coherence is established, and that the Markovian approximation is valid. This condition may nominally break down in the strong pumping regime, but we will nevertheless apply Fermi golden rule.

\section{Linewidth in thermal equilibrium}\label{sec:thermal}
In this Section, we consider the contribution to the linewidth of collective modes from intrinsic interactions in the absence of external driving. We first introduce the approximate separation of decay processes into \textit{(i)} large momentum transfer and \textit{(ii)} small momentum transfer. Then, in Sec.~\ref{subsec:largemom} we discuss resonant (on-shell) processes of type \textit{(i)}, recovering previous results in the literature \cite{bard_prb_2018,lin_prl_2013}. Finally, in Sec.~\ref{subsec:offreson} we analyze off-resonant processes allowed by broadening, of both types \textit{(i)-(ii)}. For all processes, we obtain the scaling of decay rate with mode number and temperature and support these findings with numerical simulations. 

\subsection{Different processes contributing to decays}
\begin{figure}[b] 
        \includegraphics[width =0.99 \columnwidth]{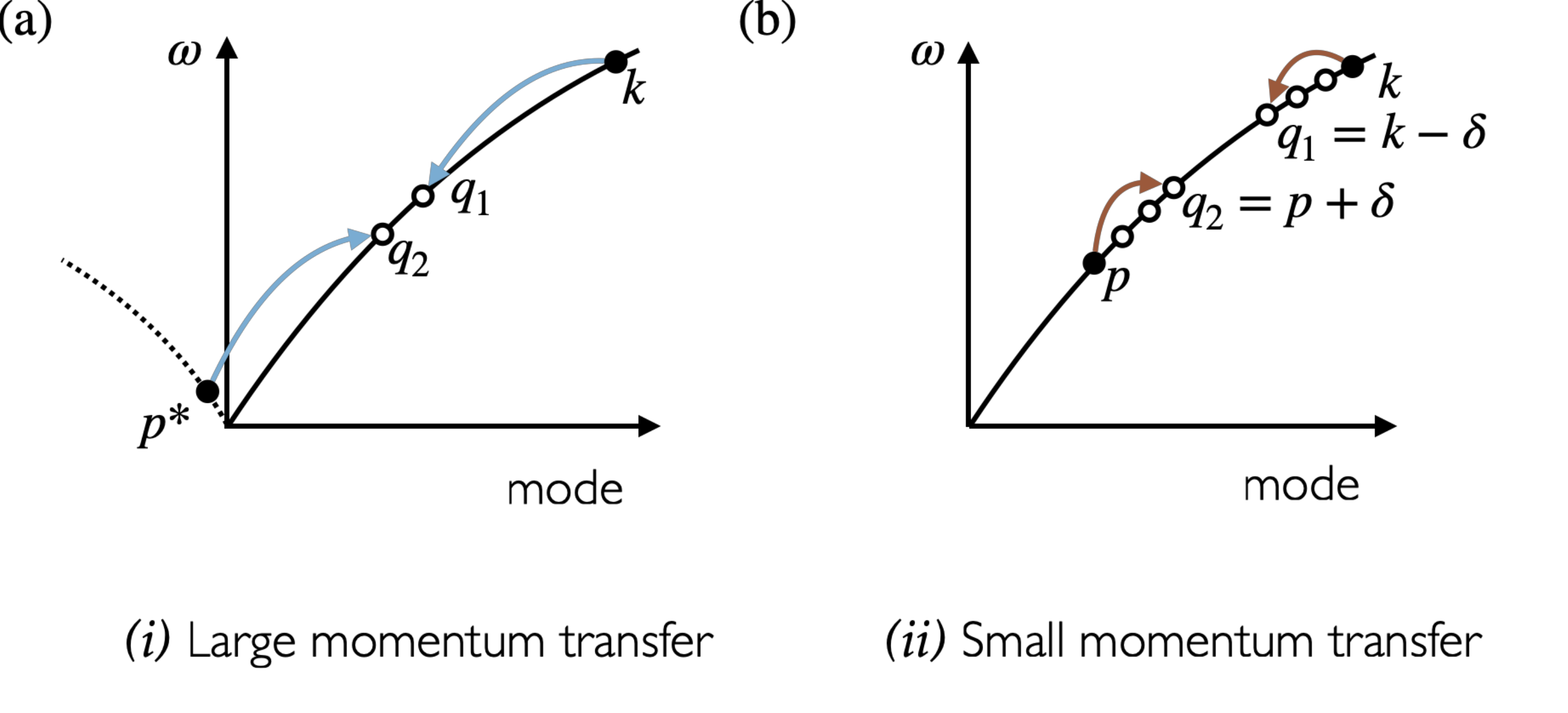}
        \caption{Schematic representation of large \textit{(i)} and small \textit{(ii)} momentum transfer decay of mode $k$. The only on-shell decay process involves momentum $p^*$ defined in the main text.}\label{fig:final4}
\end{figure}

To start, we qualitatively characterize the possible decay processes of a given mode $k\gg 1$. Energy and momentum conservations put strong constraints on the decay processes allowed with a concave dispersion relation as Eq.~\eqref{eqn:mirlindisp} in a JJ chain. In particular, previous works~\cite{lin_prl_2013,bard_prb_2018} obtained the following condition for the decay of mode $k$ into mode $q_1$ with simultaneous up-scattering of mode $p^*$ to mode $q_2$ that conserves energy and momentum: 
\begin{equation}\label{Eq:pstar}
p^*  = - \frac{3}{2}\xi k q_1q_2, 
\end{equation}
where an expansion assuming the dispersion's nonlinearity parameter $\xi\ll 1$ (see Eq.~\eqref{eqn:mirlindisp}) was performed. For realistic values of parameter $\xi$, that are typically of the order of $10^{-7}-10^{-5}$, the existence of non-trivial solutions to this condition implies that momenta $q$ are as large as possible, suggesting that $q_1\approx q_2 \approx k/2$. We denote this type of processes as \emph{large momentum transfer} and illustrate them in Fig.~\ref{fig:final4}(a).

Due to the finite linewidth of modes, processes that weakly violate the energy conservation due to extrinsic couplings are also allowed. Since the dispersion features a weak nonlinearity, the weakest energy violation is happening for \emph{small momentum transfer processes}, such as in  Fig.~\ref{fig:final4}(b), where initial modes $k\gtrsim p$ have comparable momenta, and momentum transfer $\delta = k-q_1 = q_2-p$ is small. Although the separation between large and small momentum transfer processes is not always sharp, in practice we will see that restricting the momentum transfer $\delta$ to small values of the order of few modes is typically enough to account for small momentum transfer processes. 

Finally, for both processes discussed above, we can distinguish the on-shell and off-shell contributions. The former contribution survives in the limit where the density of states becomes a delta-function. In contrast, the off-shell contribution relies on the finite broadening of modes. For large momentum transfer processes, we write:
 \begin{equation}
    \delta\kappa_k^{(i)}=\delta\kappa_k^{(i), \text{on}}+\delta\kappa_k^{(i), \text{off}}.
\end{equation}
In contrast, for the small momentum transfer, only the off-shell contribution exists ($\delta=0$ would correspond to the excluded self-decay), therefore we omit the extra label in what follows and use the notation $\delta\kappa_k^{(ii)}=\delta\kappa_k^{(ii), \text{off}}$. 

In order to support our classification of processes introduced above, we perform a numerical simulation of the total decay rate in Eq.~\eqref{eqn:currentdecays} assuming the equilibrium distribution function. In order to visualize the contributions of different  processes to the decay rate of a fixed mode $k$, we rely on momentum conservation (using the unfolded momentum notations) to solve for momentum $p=q_1+q_2-k$. The resulting two momenta $q_1$ and $q_2$ are used to label the decay process of mode $k$, and we visualize the relative importance of individual processes by plotting their contribution to the total intrinsic decay rate, Eq.~(\ref{eqn:currentdecays}).  

\begin{figure}[t] 
        \includegraphics[width =\columnwidth]{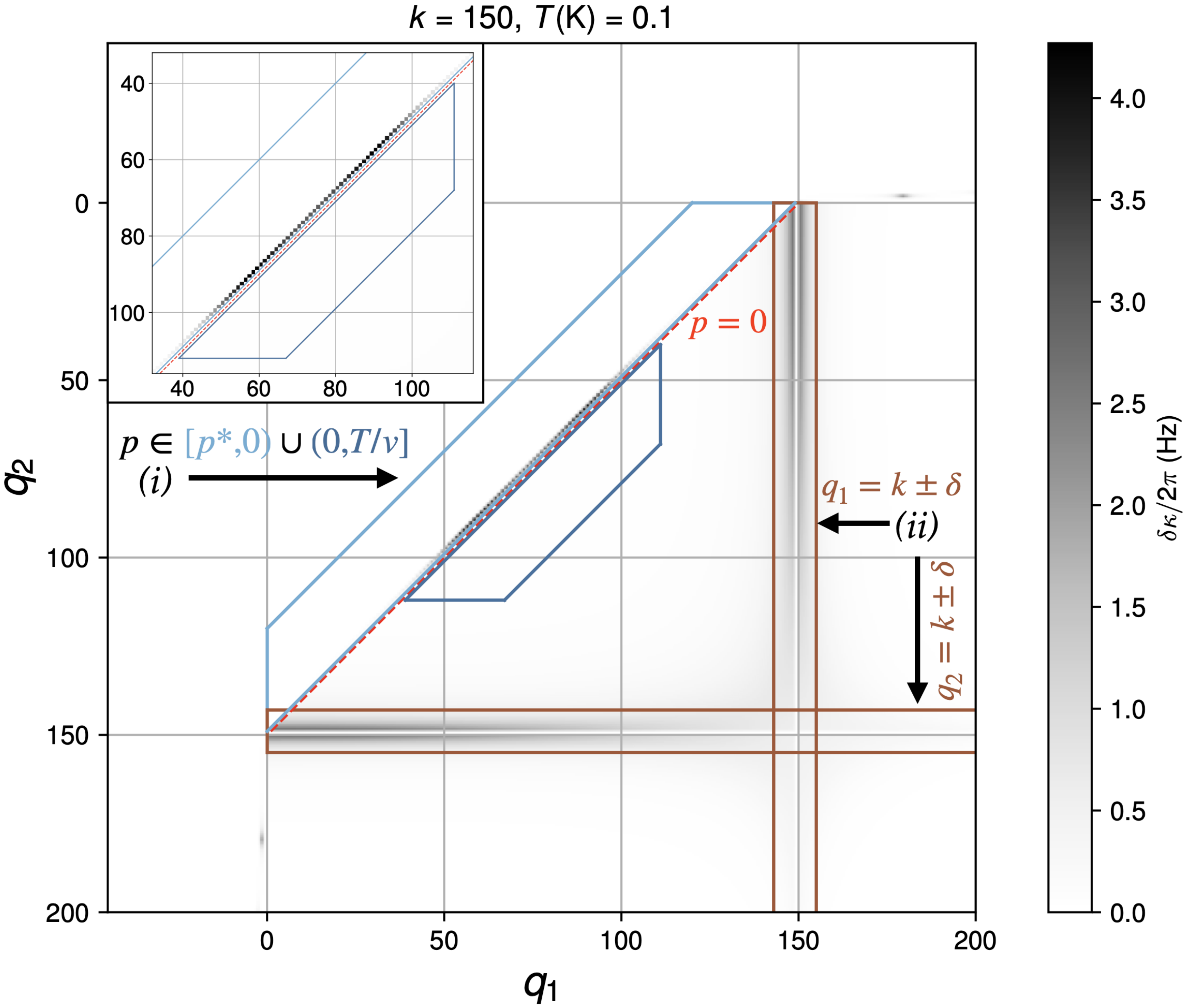}
        \caption{Decay channel plot of mode $k=150$ assuming $k,p\to q_1,q_2$, and setting $T=0.1\,\text{K}$ and $\kappa^0/(2\pi)=5\,\text{MHz}$. The darker regions correspond to participating decay channels. The colored frames highlight two relevant and distinguished types of scattering processes: with large momentum transfer \textit{(i)} (light and dark blue) and small momentum transfer \textit{(ii)} (brown), respectively. Numerically, for the resonant processes of type \textit{(i)} (see the main text), we sum $p$ over $-30\le p \le -1$, which includes $p^*$ up to $k\approx450$, and we restrict to $q_1,q_2>0$. For the off-resonant counterpart, we consider $p\in(0,T/v]$ and $k/4\le q_1,q_2\le 3k/4$, which correspond to a large momentum transfer window. Processes of type \textit{(ii)} are instead selected through $q_1,q_2>0$ and $k-\delta<q_1,q_2<k+\delta$. The choice of $\delta$ and more details on the selected channels are discussed in the main text.}\label{fig:final5}
\end{figure}

Figure~\ref{fig:final5} shows such plot for a particular choice of parameters. We selected mode $k=150$ and used the Bose-Einstein distribution function with temperature $T=0.1\, \text{K}$. Here and unless differently specified, we assume a relatively large mode broadening of $\kappa^0/(2\pi)=5\,\text{MHz}$, still smaller than inter-mode spacing.~\footnote{As we will show, a large broadening is required to enable all types of processes. Apart from that, a different choice would not alter qualitatively the results of this work.} The total decay rate of mode $k$ corresponds to the sum of all entries in Fig.~\ref{fig:final5}, with $q_1$ and $q_2$ spanning all non-zero values from $-k_\text{max}$ to $k_\text{max}$. Modes outside the plotted window in Fig.~\ref{fig:final5} make a vanishingly small contribution and are not shown in the Figure. The notable parts of the plot that have considerable weights are labeled as \textit{(i)} and \textit{(ii)}, and they match our large and small-momentum transfer processes defined above.

Indeed, the large momentum transfer process has $p^*$ small and negative, and we see a thin gray line near the $p=0$ diagonal, that is also highlighted in the inset of Fig.~\ref{fig:final5}. Here, the momentum transfer between initial and final states, $\delta=k-q_1=q_2-p$, can be of order $k/2$. The second set of notable features denoted as \textit{(ii)} corresponds to a scattering process where the momentum transfer $\delta$ is small, $q_1\approx k$ or $q_2\approx k$. Two additional weak features are visible in the top right corner and in the bottom left corner of Fig.~\ref{fig:final5}, again stemming from large momentum transfer processes, but now with mode $k$ being up-scattered. Since they are suppressed at low temperature compared to \textit{(i)}, and we will show that \textit{(ii)} dominates at higher temperature, we will not discuss these processes.

\subsection{Resonant scattering, large momentum transfer}\label{subsec:largemom}
We start from the case of resonant large momentum transfer, see Fig. \ref{fig:final4}(a). To select the contribution from these processes, we consider only terms with $p<0$ in the sum within Eq.~\eqref{eqn:currentdecays}, and replace the Lorentzian density of states by delta-function. Using the simplified expression~\eqref{eqn:boxmatrixel} for the matrix element in~\eqref{eqn:Wrate}, we obtain
\begin{multline}\label{Eq:kappa-i}
    \delta\kappa_k^{(i), \text{on}} = \frac{\pi^5 E_g^2}{16N^6}\sideset{}{'} \sum_{q_2>0,p<0}|kp(k+p-q_2)q_2|\delta(\Delta \omega)\\
    [n^0_p(1+n^0_{p+k-q_2}+n^0_{q_2})-n^0_{p+k-q_2}n^0_{q_2}],
\end{multline}
where $\Delta \omega  = \omega_{q_1}+\omega_{q_2}-\omega_{k}-\omega_{p}$ and the primed sum implies the restriction $q_1 = p+k-q_2>0$. 

A non-vanishing on-shell contribution requires the existence of mode $p^*$ as in Eq.~(\ref{Eq:pstar}), i.e.~$|p^*|$ must be larger or equal to one, resulting in the condition $k\geq \sqrt[3]{{8}/({3\xi})}$.~\footnote{Here we used $q_1=q_2=k/2$, which gives the largest possible $p^*$.} For the considered chain parameters, this yields $k \geq 150 $. In addition, the mode with index $p^*$ must be occupied by at least one excitation to considerably contribute to the decay process, suggesting that temperature should be sufficiently large. However, provided that we are typically concerned with modes $k$ in a range $\sim1-10\,\text{GHz}$, with our chain parameters the resulting $|p^*|$ is of order up to $\sim30$. Typical temperatures of order $0.1\,\text{K}$ lead to the occupation of (several) tens of modes, so this condition does not represent a considerable restriction.

Assuming that momentum $p^*$ is not too small ($|p^*|\ge\kappa^0/v$), since the expansion of $\Delta \omega$ features non-analyticity around $p=0$, we linearize the energy difference around $p\approx p^*$, obtaining $\Delta\omega \approx -2v(p-p^*)$ up to corrections of order $\xi$. Using this linearization and neglecting the occupation numbers of all modes except for $n_p$, we obtain 
\begin{multline}
    \delta\kappa_k^{(i), \text{on}} = \frac{\pi^5 E_g^2}{16N^6} \sum_{q_2=1}^{k+p} \sum_{p<0} |kp(k+p-q_2)q_2|
    \\
    \times \delta(2v (p-p^*))n_p.
\end{multline}
We note, that replacing the broadened density of states by a delta-function to obtain the on-shell contribution is justified if at least a few modes lie within the Lorentzian width, which is $2\kappa^0/v$ after linearization. This requires the inter-mode spacing given by $v$ to be smaller than the mode width $\kappa^0$, $\kappa^0>v/2$. This condition practically does not hold, so the approximation is not valid. Nevertheless, let us give for completeness a closed expression for $ \delta\kappa_k^{(i), \text{on}}$, that is obtained from approximately replacing sums by integrals: 
\begin{equation}
    \delta\kappa_k^{(i), \text{on}} = \frac{\pi^5E_g^2}{192v^2N^6}Tk^4= \frac{\pi^3E_g}{384N^4E_J}Tk^4. \label{eqn:mirlintheory}
\end{equation}
This expression matches the result of Ref.~\cite{bard_prb_2018}, provided that periodic boundary conditions are replaced by open boundaries and notations are matched. A violation of $\kappa^0>v/2$ would affects in particular the temperature scaling, which is expected to be higher than that predicted by Eq.~\eqref{eqn:mirlintheory}.

The obtained scaling of the intrinsic decay rate from this resonant large-momentum scattering~(\ref{eqn:mirlintheory}) can be reproduced in an intuitive way by taking into account the existence of a resonance at small momentum $p=p^*$ and using the occupation number $n_{p}\approx T/(vp)$. The scattering matrix element contributes a term $\propto k^3 p$, as all other momenta $q_1\approx q_2$ are of order $k$, and the remaining sum over $q_2$ gives another factor of $k$. This in total gives a scaling $\delta\kappa_k^{(i), \text{on}} \propto |k^4 p^*|\cdot T/(v|p^*|) \propto (T/v) k^4$ and reproduces the power laws obtained above and in Ref.~\cite{bard_prb_2018}. However, as already discussed, the main obstacle for implementing this process is the discrete nature of the chain and the level spacing between modes exceeding the linewidth of individual modes. 

\begin{figure*}[t] 
        \includegraphics[width=\textwidth]{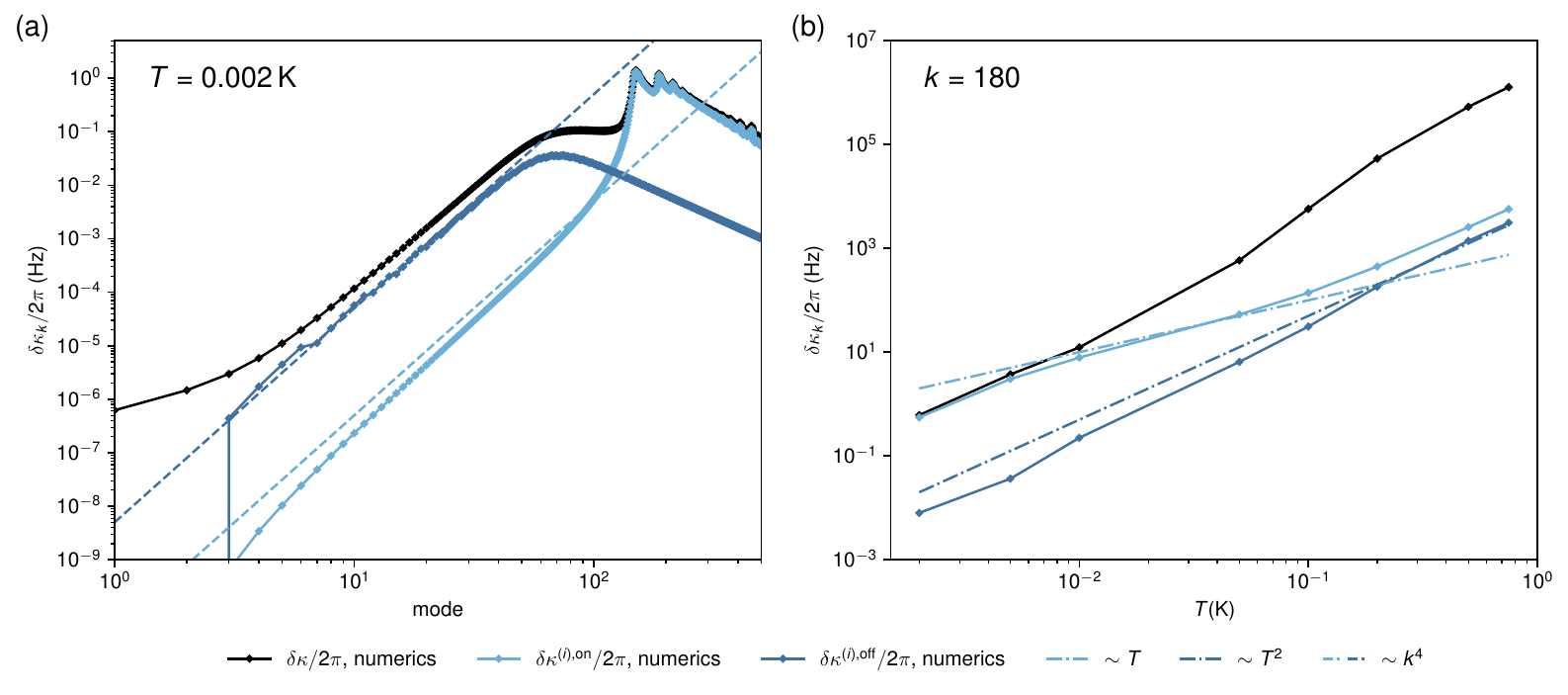}
        \caption{Analysis of decays arising from process \textit{(i)}, involving a large momentum transfer between the two initial and the two final states. (a) Excess linewidth $\delta\kappa/2\pi,\delta\kappa^{(i),\text{off}}/2\pi,\delta\kappa^{(i),\text{on}}/2\pi$ as a function of mode number, for $T=0.002\,\text{K}$. The on-shell contributions $\delta\kappa^{(i),\text{on}}/2\pi$ strongly underestimate the total decay rate $\delta\kappa/2\pi$, although well-obeying the predicted power law $\sim k^4$ scaling with mode number. (b) The same quantities for fixed mode $k=180$, as a function of temperature. Above $T=0.005\,\text{K}$, small momentum transfer processes start to contribute to $\delta\kappa$, causing the deviation between the full decay rate and its estimate from large momentum transfer processes only. The contribution from off-resonant large momentum processes, $\delta\kappa^{(i),\text{off}}/2\pi$, follows the expected $\sim T^2$ scaling, while the contribution from resonant processes shows a weaker scaling although not exactly $\sim T$ (not all conditions for such scaling are fulfilled).}\label{fig:final6}
\end{figure*}

\subsection{Off-resonant scattering}\label{subsec:offreson}
We proceed by considering off-resonant decays. Although it is practically impossible to observe the resonant large-momentum scattering, the finite linewidth $\kappa^0$ facilitates large-momentum transfer off-shell contributions that we consider first. These processes are most relevant in the low-temperature regime. After this, we continue with the estimate of small momentum transfer processes that have only off-shell contribution, and dominate at larger temperatures. 

\subsubsection{Large momentum transfer}
The off-shell contributions to large momentum transfer decay processes are estimated from Eq.~(\ref{Eq:kappa-i}), by restricting the sum over $p$ to positive values $p\in\left(1,{T}/{v}\right)$ and restoring the Lorentzian form of the density of states. To give an analytical estimate, we consider a decaying mode $k$ with energy much larger than temperature. Also, we assume $k\ll \sqrt[3]{8\kappa^0/(3v\xi)}\approx 61$, which comes from $\kappa^0/v\gg |p^*|$ and is complementary to our last requirement for resonant processes. This condition places the Lorentzian $\delta_\gamma(\Delta\omega)$ centered in $p^*$ close to $p=0$, where  $\delta_\gamma(\Delta\omega)$ deviates from a Lorentzian, becoming roughly constant in $p$. This implies that channels from the positive branch $p>0$ acquire importance, and comes from the non-analytical dependence of the linearized $\Delta\omega= \omega_{q_1}+\omega_{q_2}-\omega_{k}-\omega_{p}$ on $p$ when $p$ changes sign. Indeed, for negative values of $p$ above we used the expansion $\Delta\omega(p)\approx \Delta\omega(0) -2vp$ to derive $\delta\kappa_k^{(i), \text{on}}$. In contrast, for positive values of $p$, the expansion reads $\Delta\omega(p)\approx\Delta\omega(0)+3k^2v\xi p/4$, where the presence of the small parameter $\xi$ that controls the dispersion nonlinearity leads to a much slower energy change with $p$. As a result, we have $|\Delta\omega(p)-\Delta\omega(0)|\ll\gamma$ for all values of $p$ where at least one excitation is present, $n_p\geq 1$. Calculating the sums, we get
\begin{equation}
    \delta\kappa_k^{(i), \text{off}} = \frac{\pi^3E_g}{384N^4E_J}\frac{T^2}{\kappa^0}k^4.\label{eqn:mirlinofftheory}
\end{equation}
The scaling of this off-resonant contribution is intuitively obtained in the same way as before, however replacing the density of states by a factor of $1/\gamma \propto 1/\kappa_0$ and restricting the sum over $p$ to values $p\leq T/v$, thereby obtaining another factor of temperature.

Figure~\ref{fig:final6} presents the numerical results for resonant and off-resonant scattering with large momentum transfer together with the total decay $\delta\kappa_k$. The latter is obtained by summing over all decay channels in Eq.~\eqref{eqn:currentdecays}. In Fig.~\ref{fig:final6}(a) we plot decay rates as a function of mode number at a low temperature $T=0.002\,\text{K}$. The resonant decay rate, $\delta\kappa_k^{(i), \text{on}}$ (light blue) is much smaller compared to the total decay rate $\delta\kappa_k$ (black) until $k\approx150$. This is in agreement with the our criteria for the physical existence of resonant scattering mode $p^*\geq 1$, see discussion below Eq.~\eqref{Eq:kappa-i}. For lower modes, $k\leq 60$, $\delta\kappa_k$ is instead well approximated by $\delta\kappa_k^{(i), \text{off}}$ (dark blue). Numerically, the latter is selected by summing over $p\in\left(0,{T}/{v}\right)$ and imposing a constraint $k/4\le q_1,q_2\le3k/4$. 

We notice that although $\delta\kappa_k\approx \delta\kappa_k^{(i)}$, the excess linewidth is much smaller than the bare one, $\kappa^0$. Such small changes in the linewidth are challenging to detect experimentally, therefore the scalings predicted here, $\delta\kappa\sim T^2 k^4$ or $\delta\kappa\sim T k^4$, will be hard if not impossible to verify experimentally. In Fig.~\ref{fig:final6}(b) we plot the excess linewidth as a function of temperature for a fixed mode number $k=180$. Above $T\approx 0.001\,\text{K}$, we observe a change in the temperature scaling of $\delta\kappa_k$ from $\sim T^2$ to a higher power law. At the same time the difference between total excess linewidth and the contribution from large-momentum transfer processes increases by an order of magnitude. This indicates that additional decay channels beyond large momentum transfer, \textit{(i)}, appear. These are small momentum transfer processes, which we discuss next. 

\subsubsection{Small momentum transfer}\label{subsec:smallmom}
To understand the discrepancy between the full contribution to the intrinsic scattering rate, $\delta\kappa_k$, and the contribution from large momentum transfer processes considered above, $\delta\kappa_k^{(i)}$, we consider now small momentum transfer processes \textit{(ii)} from Fig.~\ref{fig:final4}(b). To estimate analytically their contribution, as above we restrict to decays where $q_1,q_2>0$ and $0<p\le T/v$. However now, importantly, we consider transferred momenta up to a certain cutoff, $|\delta|=|k-q_1|=|q_2-p|\le\delta_\text{max}$, and assume that this cutoff is much smaller than the number of mode for which we consider decay processes, $k\gg\delta_\text{max}$. The choice of $\delta_\text{max}$, and thus the number of neighboring modes that contribute, mainly depends on the bare linewidth and the curvature of the spectrum, as we discuss below, but turns out to be of the order of few modes. 

We first linearize the energy difference $\Delta\omega = \omega_{k+\delta}+\omega_{p-\delta}-\omega_{k}-\omega_{p}$  around $\delta=0$, which is its minimum corresponding to self-decay, 
\begin{equation}\label{Eq:deltaomega}
    \Delta\omega \approx -3v \xi(k^2 - p^2)\delta.
\end{equation}
Self-decay processes, corresponding to $\Delta\omega$ being exactly zero, are excluded, explaining the absence of resonant contribution to small momentum transfer decays. Substituting this into Eq.~\eqref{eqn:currentdecays} gives
\begin{multline} \label{Eq:kappa-ii}
    \delta\kappa_k^{(ii)} =\frac{\pi^5 E_g^2k}{16N^6}\sideset{}{'}\sum_{\delta=-\delta_\text{max}}^{\delta_\text{max}}\sum_{p=1}^{T/v} (p-\delta)(k+\delta)p
    \\
    \times \delta_{\gamma}(3v \xi(k^2 - p^2 ) \delta)
n_p(1+n_{p-\delta}),
\end{multline}
where we retained only the factor $n_p$ and simplified $n_pn_{k+\delta}-n_{p-\delta}n_{k+\delta}\approx0$. This is justified as long as $p-\delta\approx p$. The primed sum over $\delta$ denotes that we exclude self-decay, $\delta\neq0$. The remaining Bose occupation part is therefore $n_p(1+ n_{p-\delta}) \approx T/(vp)\{1+T/[v(p-\delta)]\} $. 

\begin{figure*}[tb] 
        \includegraphics[width=\textwidth]{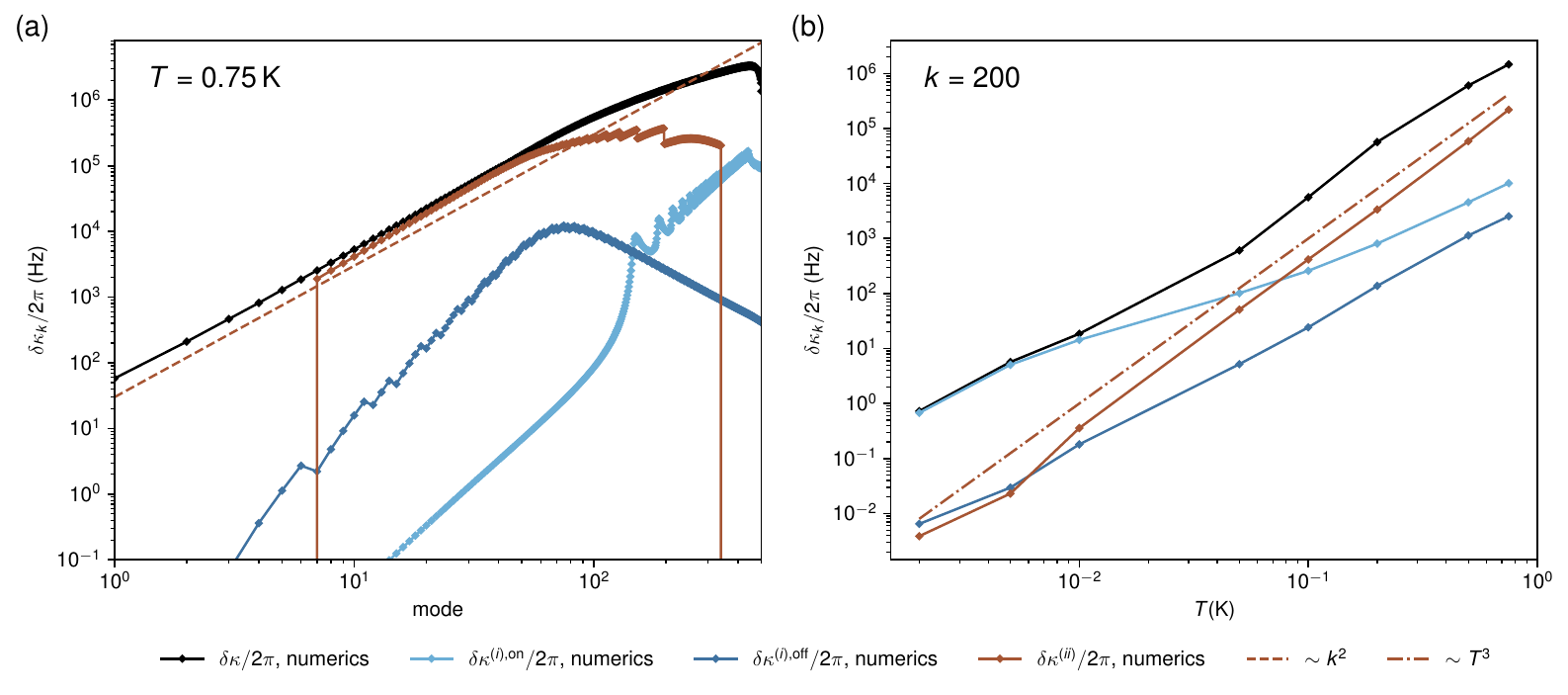}
        \caption{Contribution of small momentum transfer processes \textit{(ii)} to decay. In both panels, we use $\delta_{\text{max},k}$ calculated from Eq.~\eqref{Eq:delta-max}, although this is not correct for the lowest temperatures and mode numbers. (a) Total decay rate $\delta\kappa/2\pi$ and individual contributions $\delta\kappa^{(i)}/2\pi$ and $\delta\kappa^{(ii)}/2\pi$ as a function of mode number for $T=0.75\,\text{K}$ 
        shows that small momentum transfer processes~\textit{(ii)}  dominate over large momentum transfer~\textit{(i)}. The small momentum transfer processes follow the predicted power law scaling $\sim k^2$ with mode number. Compared to $\delta\kappa^{(ii)}$, the total decay maintains the same scaling for a broader range of mode numbers. We emphasize that such scaling is entirely due to small momentum transfer processes, as can be checked by summing numerical contributions up to a $k$-independent $\delta_{\text{max}}$ of the order of few modes. The discrepancy originates from our simplified approach, which considers the lowest $\delta_{\text{max},k}$ over all $p$-range, resulting in suppressed contributions. (b) Dependence of decay rate of the mode with fixed number $k=200$ as a function of temperature. Around $T=0.05\,\text{K}$, we observe a crossover in $\delta\kappa$ between a lower exponent in $T$ and $\sim T^3$, and the prevalence of processes of type \textit{(ii)} over \textit{(i)}.
        }\label{fig:final7}
\end{figure*}

In order to estimate this expression, we restrict the summation over momentum transfer value $\delta$ to the region where the density of states is the largest and approximately constant, therefore replacing the Lorentzian by a box $\delta_{\gamma}(\Delta\omega)\approx1/(\pi\gamma)$ with appropriate width. The latter can be determined from the condition of $\Delta\omega$ in Eq.~(\ref{Eq:deltaomega}) not exceeding $\Delta\omega \approx \gamma = 2\kappa_0$. This gives us the estimate for the upper boundary of the summation
\begin{equation}\label{Eq:delta-max}
    \delta_{\text{max},k}\approx \frac{\gamma}{3v\xi \mathop{\rm max}_{p\in\{1,T/v\}}|k^2-p^2|}.
\end{equation}
For any $|\delta|<\delta_{\text{max},k}$, the approximation $\delta_{\gamma}(\Delta\omega)\approx1/(\pi\gamma)$ remains valid across the entire momentum range $p\in[1,T/v]$. We observe that if the curvature is too large, $\delta_{\text{max},k}\lesssim 1$ and this scattering contribution is suppressed. Likewise, the decrease of $\delta_{\text{max},k}$ with the maximum separation between $k$ and allowed momenta $p$, also reflects curvature effects. Contrarily, if either temperature or mode number are too low, where the lowest and almost linear part of the spectrum is involved, $\delta_\text{max}$ from Eq.~\eqref{Eq:delta-max} might be comparable or larger than $k$ and $p$. In this case, the approximations we did so far fail, and one might include further constraints or investigate just the decay into the nearest modes, $\delta=1$. However we are primarily interested in high temperatures, where $\delta\kappa_k^{(ii)}$ dominates. Here $\delta_\text{max}$ is typically of the order of a few modes (typically, less than 10), therefore these processes involve a small momentum transfer between initial and final states as anticipated.

Using the approximated $\delta_{\gamma}(\Delta\omega)$, and assuming $k+\delta\approx k$ and $n_{p-\delta}\approx n_p$, we simplify Eq.~(\ref{Eq:kappa-ii}) as 
\begin{multline}
    \delta\kappa_k^{(ii)}\approx\frac{\pi^4 E_g^2k^2}{16N^6\gamma} \frac{T}{v}\sum_{\delta=-\delta_{\text{max},k}}^{\delta_{\text{max},k}}\sum_{p=1}^{T/v}\left(p+\frac{T}{v}\right)\\\approx\frac{\pi^4 E_g^2k^2}{16N^6\gamma} \frac{3T^3}{v^3}\delta_{\text{max},k}.\label{eqn:nearbytheory}
\end{multline}
Let us unpack the resulting scalings depending if $k$ is larger or smaller compared to $k^*=\sqrt{(1 + T^2/v^2)/2}\approx T/(\sqrt{2}v)$. When $k\ll T/v$, we obtain the scaling of excess linewidth $\delta\kappa_k^{(ii)}\propto {k^2(T/v)^3}/[{(T/v)^2 - k^2}]$. In the opposite case, we get $\delta\kappa_k^{(ii)}\propto k^2(T/v)^3/[{k^2 - 1}]$. Simplifying this to the leading order terms, we get
\begin{equation} \label{Eq:scaling-ii}
\delta\kappa_k^{(ii)}\approx
\frac{\pi^4E_g^2}{16N^6\xi}
\begin{cases}
\frac{k^2T}{v},
& k \ll T/v \\
\frac{T^3}{v^3},
& k \gg T/v
\end{cases},
\end{equation}
where we see that $\delta\kappa_k^{(ii)}$ becomes $k$-independent for sufficiently large $k$. Although these are two limiting cases, in practice we see scaling of excess linewidth as $\propto k^2$ for fixed $T$ and $\propto T^3$ for fixed $k$ for a broad range of mode numbers and temperatures. The obtained scalings $\delta\kappa_k^{(ii)}$ in Eq.~(\ref{Eq:scaling-ii}) as a function of mode number and temperature allow for a simple and intuitive explanation. To this aim, let us assume for simplicity that $\delta_{\text{max}}$ is fixed for all $k$. The matrix element square is proportional to the product of four mode numbers, and for process \textit{(ii)} we can approximate such product as $k^2p^2$. Then, the remaining sum over $p$ up to temperature reproduces the same scaling $\delta\kappa_k^{(ii)} \propto k^2 T^3/v^3$ in agreement with Eq.~(\ref{eqn:nearbytheory}).

In Figure~\ref{fig:final7}(a), we show decay rates as a function of mode numbers at a high temperature $T=0.75\,\text{K}$, where the decay channel plot shows the dominant contribution coming from small momentum transfer processes. We find that $\delta\kappa_k^{(ii)}$ (brown) captures well the total decay rate $\delta\kappa_k$ (black), and shows a characteristic increase with the square of mode number $k^2$. If $\Delta\omega$ is larger than $\gamma$, and $\delta_{\gamma}(\Delta\omega)$ cannot be simplified, then for sufficiently large $k$ we expect a decrease of decay rate as $\sim k^{-2}$ with mode number $k$, which is indeed observed in simulations (see Fig.~\ref{fig:(ii)quant2} in Appendix \ref{appendixF}).
Finally, Fig.~\ref{fig:final7}(b) shows the decay rate of mode $k=200$ as a function of temperature. Around a temperature of $0.08\,\text{K}$ we see a crossover from large momentum transfer processes \textit{(i)} to small momentum transfer processes \textit{(ii)} that dominate $\delta\kappa_k$ (the same crossover is responsible for the deviation observed in Fig.~\ref{fig:final6}(b)). Consequently, the total decay rate follows the expected power law $\sim T^3$ increase with temperature, in agreement with Eq.~(\ref{eqn:nearbytheory}).

\section{External pumping and NESS}\label{sec:noneq}
In this Section we move into the non-equilibrium setting, achieved by pumping a photon flux into the JJ chain. First, in Section~\ref{subsec:simulation} we discuss the details of the numerical simulation we use to extract the NESS of the system and the excess linewidth of different modes. In Section~\ref{subsec:vnadrive} we consider the regime where the system is not too far from equilibrium, and demonstrate how large momentum transfer processes can be enhanced and observed. In Section~\ref{subsec:narrowing} we explore linewidth narrowing due to the presence of weak driving. Finally, in Sec.~\ref{subsec:strongdrive} we consider the strongly driven regime. We discuss the scaling of excess linewidth with the flux intensity, and the crossover to the regime where the memory of the pumping configuration is lost.

\subsection{Numerical simulation details}\label{subsec:simulation}
In order to obtain the occupation numbers of individual modes in the NESS, denoted as $n^0_k$, we use the time-evolution of the kinetic equation~(\ref{eqn:boltzkineq}) with a na\"ive first order explicit time propagation scheme. To calculate the collision integral we use the same code as in Ref.~\cite{bubis_sciadv_2026}. We keep a mode-independent bare linewidth $\kappa^0$ and neglect Kerr shifts. Energy conservation is therefore enforced via the Lorentzian $\delta_\gamma(\Delta \omega)$ in Eq.~\eqref{eqn:dos}, with $\gamma = 2\kappa^0$. The excess linewidth $\delta\kappa_k$, accounting for intrinsic nonlinear scattering, is then computed from Eq.~\eqref{eqn:currentdecays}. In the strong-drive regime, $\delta\kappa_k$ can become comparable to $\kappa^0$ and thus modify scattering probabilities. In principle, the linewidth should be updated self-consistently throughout the evolution; see Ref.~\cite{bubis_sciadv_2026} for a simplified implementation. Since applying the same approximate scheme here does not alter the main conclusions drawn in this work, unless the driving is very strong (see Sec.~\ref{subsec:strongdrive}),  we do not include it. 

\begin{figure*}[tb] 
\centering
        \includegraphics[width=\textwidth]{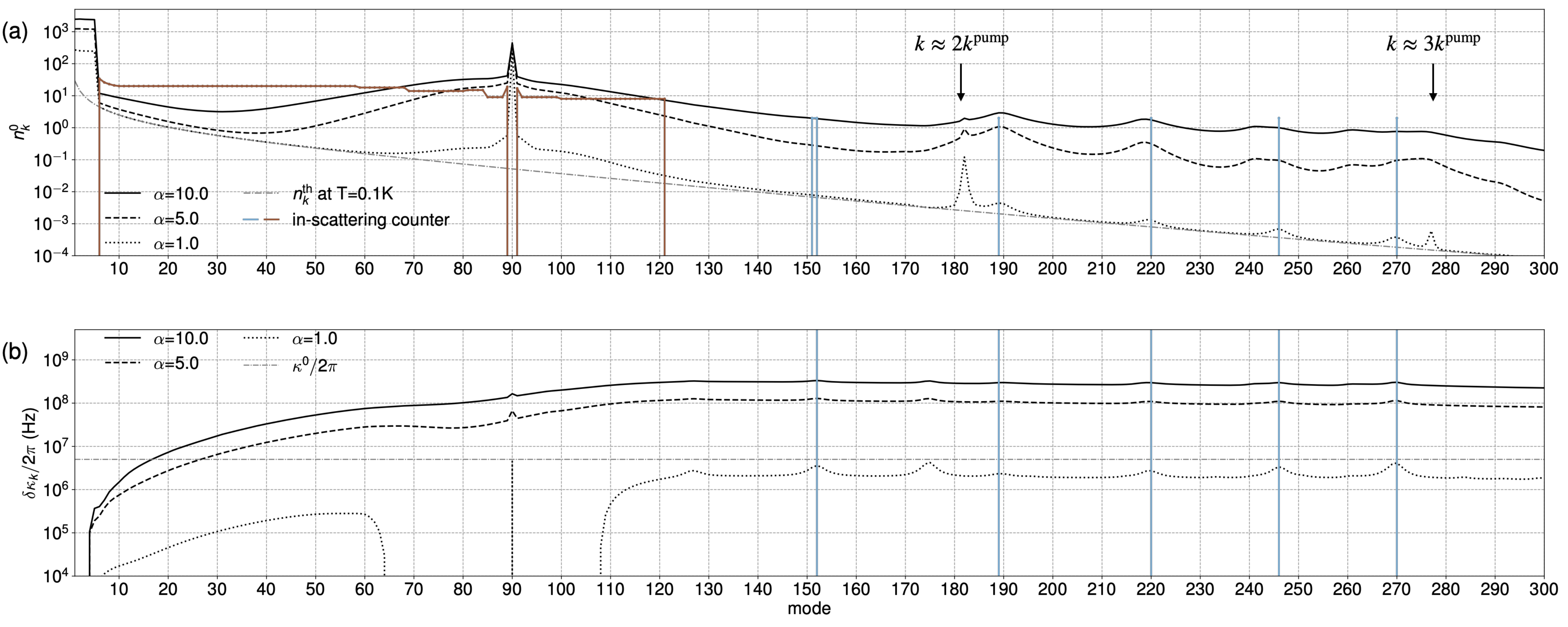}
        \caption{(a) NESS resulting from pumping modes ${\mathbb K} =\{1,2,\ldots,5\}$ and $k^\text{pump}=90$ with different flux intensities, $\alpha$, as specified in the legend. The superimposed histogram indicates the in-scattered photons from collisions where two modes out of $p,q_1,q_2$ are in the set of pumped states, and $|\Delta\omega|<\kappa^0/2$. Enhanced population is expected in the vicinity of pumped modes (brown), as a result of small-momentum transfer scattering \textit{(ii)}, and in isolated modes (blue), which we identify as large-momentum transfer on-shell processes \textit{(i)}. These peaks are visible at low pump intensities and get smoothed out at sufficiently high pump power. Additional peaks which survive at high intensity are observed at multiples of pumped mode, $2 k^\text{pump}$ and $3 k^\text{pump}$. (b) Excess linewidth as a function of mode number for the three NESS in (a). The horizontal line indicates the bare linewidth $\kappa^0$. We find several peaks, most of which at the same location of enhanced occupation in (a). Notably, a broad range of modes around the isolated pumped mode $k^\text{pump}$ exhibits a negative excess linewidth.}\label{fig:peaks}
\end{figure*}

For the photon flux, we choose parameters that are comparable to those reported in the experimental work~\cite{bubis_sciadv_2026}.
The coupling to terminals is taken to be $\kappa_\text{ex}/(2\pi)=2\,\text{MHz}$,  and we define the photon flux as
\begin{equation}
n_k^\text{flux} = \alpha {n}_k^\text{b}
\  \
\text{with} \  n_k^\text{b}=600\,\frac{\text{ph}}{\text{Hz}\cdot\text{s}}
\end{equation}
for any pumped mode, where $\alpha$ is the dimensionless flux intensity. From Eq.~\eqref{eqn:boltzkineq}, this choice of parameters corresponds to an injection of $\approx \alpha\cdot10^9\,\text{ph/s}$. In the following, we will consider two pumping configurations, possibly combined: the pumping of a low-energy set of modes, denoted as $\mathbb{K}$, and the pumping of a single higher energy mode, denoted as $k^\text{pump}$, $k^\text{pump}\notin \mathbb{K}$. For simplicity, both setups are incorporated in the kinetic equation through $n_k^\text{flux}$. Experimentally, the first scheme is realized with a broadband pumping with constant power spectral density within the selected frequency range; the second one, with a single tone at the frequency matching the desired mode. 
While, in principle, the experimental setup allows for a great flexibility in the choice of the pumped modes, we will consider the natural situation when mostly low-energy modes with the smallest $k=1,2,\ldots$ are pumped. Indeed, the system is subject to thermal noise at the lowest frequencies, and such pumping configuration mimics enhanced noise. It will also be shown to facilitate resonant, large momentum transfer scattering.

\subsection{Enhancing resonant scattering with weak pumping}\label{subsec:vnadrive}
We start by considering the pumping of the first 5 modes and mode $k^\text{pump}=90$. In Fig.~\ref{fig:peaks}(a), we show the steady state occupation numbers resulting from solving the kinetic equation for different pump intensities, $\alpha=1,5,10$. While the analytic solution for the NESS is not possible in this situation, we can qualitatively understand the results of the numerical simulation using the intuition for different scattering processes we developed in Section~\ref{sec:thermal}.

For the smallest intensity $\alpha=1$, the resulting $n^0_k$ in Fig.~\ref{fig:peaks} deviates from the Bose-Einstein distribution function at base temperature, for small $k\leq k^\text{pump}$, only near the pumped modes. This deviation can be attributed to small momentum transfer events that are off-resonant. Notably, at larger $k\geq 2 k^\text{pump}$ we see an additional weak enhancement of $n^0_k$ compared to thermal distribution. Two of these peaks, labeled as $\approx 2 k^\text{pump}$ and  $\approx 3 k^\text{pump}$ peaks, emerge from up-scattering of several excitations from the only driven high-energy mode. In particular, the two-excitation up-scattering process works as $k^\text{pump}, k^\text{pump} \to 2 k^\text{pump}+\delta,-\delta$, where the value of $\delta$ is small. This process is not resonant, but due to the strong pumping of mode $k^\text{pump}$, it causes a significant increase of occupation. In addition to these peaks at approximately multiple momenta of $k^\text{pump}$, we observe several further features.

The additional peaks highlighted by blue lines in Fig.~\ref{fig:peaks}(a) become more prominent as intensity is increased to $\alpha=5$, however for even larger intensity $\alpha=10$ they become less visible as the distribution function strongly deviates from the Bose-Einstein distribution function everywhere. The location of blue lines is obtained independently of the solution of kinetic equation, to show resonant scattering processes. To this end, we consider all 2-into-2 scattering events $k,p \to q_1, q_2$, where two modes out of  $\{p, q_1, q_2\}$ belong to the set of driven modes $\mathbb K\cup\{k^\text{pump}\}$. Among those combinations, we highlight mode $k$, provided the energy difference is zero within half a mode linewidth $\kappa^0/2$ to allow for a very small displace from resonance, $|\omega(k)+\omega(p)-\omega(q_1)-\omega(q_2)|\leq {\kappa^0}/2$. For $\omega(k)$ we use the full expression in Eq.~(\ref{eqn:exactdisp}). For the specific pumping configuration considered here, we get resonant in-scattering into modes ${k=151,152,189,220,246,270}$ where blue peaks are shown. We find that these peaks correspond to resonant scattering with large momentum transfer, where $k^\text{pump}$ scatters off the negative component of a pumped mode in $\mathbb{K}$, $p^*=-1,-2,-3,-4,-5$ (see the schematic of Fig. \ref{fig:final4}(a)). These agree well with the deviation of $n^0_k$ from the thermal equilibrium value, except for the first two where no enhancement is noticeable. The absence of the enhancement of $n^0_k$ for smallest values of $k$ in the set of resonant modes may be attributed to the matrix element, that increases with mode numbers. 

In addition to the enhancement of mode occupation, we also observe an increase of linewidth for resonant modes, visible as peaks in Fig.~\ref{fig:peaks}(b) that shows the excess linewidth, calculated using Eq.~\eqref{eqn:currentdecays}. Although these peaks are found only in a certain window of pumping intensities, here the increase of the linewidth is within the experimentally detectable range. Therefore, such driving setup may be used to experimentally detect the resonant scattering process specified by Eq.~(\ref{Eq:pstar}).  More interestingly, a  range of modes in the vicinity of the isolated driven mode $k^\text{pump}$, and the first few modes with $k\le3$, exhibit a linewidth narrowing, manifested as negative contribution to the linewidth, which will be discussed in Sec.~\ref{subsec:narrowing} below. 

\begin{figure*}[t] 
\centering
        \includegraphics[width=\textwidth]{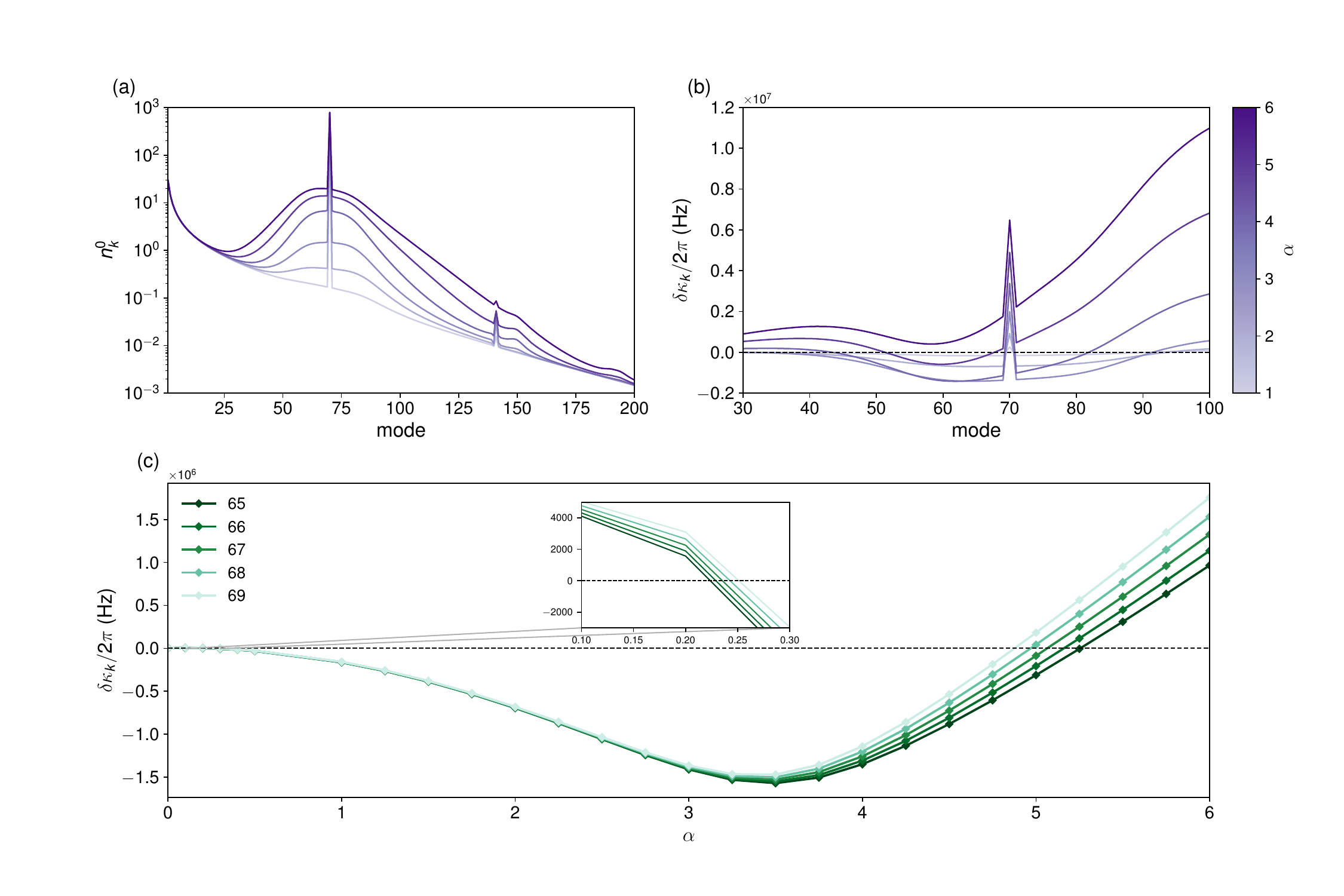}
        \caption{(a) Evolution of distribution function upon increasing pumping intensity into mode $k^\text{pump}= 70$ for $\alpha=1,2,3,4,5,6$. (b) The range where $\delta \kappa_k<0$ is first appearing upon increasing intensity, and then shrinks asymmetrically with respect to the pumped mode as discussed in the main text. Values of $\alpha$ are as in (b). (c) Evolution of $\delta \kappa_k<0$ as a function of the driving strength, for the first five modes with $k<k^\text{pump}$. The inset shows the first change in sign of $\delta\kappa_k$.}\label{fig:narrow}
\end{figure*}

In summary, the NESS in the regime of weak pumping contains signatures of resonant up-scattering of  excitations, that can lead to a non-monotonic dependence of the distribution function and mode excess linewidth on energy or mode number. We find that the conclusions obtained above for a particular pumping configuration are sufficiently general, and do not change much if one alters the number of low-energy pumped modes and the isolated pumped mode, as well as their relative pumping intensity, while maintaining the overall configuration where low several low-energy modes and a particular high-energy mode is driven.  The observed non-monotonic behavior of occupation numbers and excess linewidth is somewhat fragile, as upon increasing the intensity of photon flux, the clearly visible peaks at locations corresponding to resonant scattering tend to disappear. Nevertheless, we believe that this non-equilibrium distribution function can be probed in future experiments.

\subsection{Linewidth narrowing}\label{subsec:narrowing}
In this Section we explain the linewidth narrowing observed in Fig.~\ref{fig:peaks}(b) in terms of the non-trivial flux of excitations in the NESS. In particular, the thermal equilibrium solution of kinetic equation is obeying the detailed balance condition, prohibiting non-trivial fluxes of excitations in the steady state. The NESS induced by pumping violates the detailed balance, and some modes may experience in-flux of excitations due to intrinsic scattering processes. Below we show that upon a certain condition, such in-flux may lead to a net negative contribution to the intrinsic linewidth. 

In order to provide analytic estimates, we consider a simpler setup where only a single high-energy mode $k^\text{pump}$ is driven.  Figure~\ref{fig:narrow}(a) shows the evolution of the steady state distribution function upon increasing the pumping intensity, with base temperature set to $0.1$\,K. The negative value of $\delta \kappa_k$ is observed for modes adjacent to the pumped mode, moreover it disappears upon increasing driving strength as is shown in Fig.~\ref{fig:narrow}(b). The value of intensity where $\delta \kappa_k$ first becomes negative can be analytically estimated by comparing in- and out-scattering of a mode with a mode number $k$ within the interval $k^\text{pump}\pm\delta_\text{max}$ (see Sec.~\ref{subsec:smallmom} for a discussion of $\delta_\text{max}$ at thermal equilibrium), $k\neq k^\text{pump}$. Taking into account only small momentum off-resonant scattering, we obtain a rough analytical estimate of the intensity when in-scattering prevails over out-scattering: 
\begin{equation}\label{Eq:alpha1}
    \alpha \approx 
    \frac{2\delta_\text{max}\kappa^0}{\kappa_\text{ex}n_{k^\text{pump}}^\text{flux}}.
\end{equation}
From here we expect that the first change in sign of the excess linewidth $\delta\kappa$ occurs at approximately the same intensity for all modes closest to the pumped mode. This is consistent with the behavior observed in Figure~\ref{fig:narrow}(c), where we focus on the first five modes lower than $k^\text{pump}$. For the first five modes above  $k^\text{pump}$, the first crossover appears at slightly larger intensity ($0.4\lesssim\alpha\lesssim0.55$), and the second at slightly smaller intensity ($3.75\lesssim\alpha\lesssim4$).

Upon further increase of $\alpha$, the range of modes with negative $\delta\kappa_k$ shrinks again. This value of $\alpha$ is harder to estimate, as it emerges from the re-balancing of in- and out-scattering contributions to the given mode $k$, but now in a regime where the distribution function cannot be easily expressed analytically. While the mode receives in-scattering photons from the pumped mode ($2k^\text{pump}\to k^\text{pump}+ \delta,k^\text{pump}- \delta$), it also out-scatters excitations in modes further away from that. When such out-scattering processes again start to dominate over in-scattering, $\delta \kappa_k$ changes sign again and becomes positive. We notice that the range of modes experiencing a linewidth narrowing shrinks asymmetrically for modes lower/higher than $k^\text{pump}$. 

This can be understood with the help of Eq.~\eqref{eqn:currentdecays}. In relevant processes, one of the three momenta besides $k$ coincides with $k^\text{pump}$. Thinking about small momentum transfer processes, the remaining two will be in the vicinity of $k$ and $k^\text{pump}$, respectively. The smaller is $n_{q_1}n_{q_2}$, the larger the positive contribution is and the earlier the change in sign occurs. This happens when a higher mode $k>k^\text{pump}$ rather than a lower mode $k<k^\text{pump}$ is involved. This is consistent with Fig.~\ref{fig:narrow}: as the pump intensity increases, more population migrates to the tail of the distribution, and the range where $\delta\kappa_k<0$ gets more asymmetric. Lastly, we notice that this double change of sign in $\delta \kappa_k$ is happening only for modes near $k^\text{pump}$. In particular, our data in Fig.~\ref{fig:peaks}(b) did not reveal any change of sign of $\delta \kappa_k$ for high-lying modes, despite they receive a flux of incoming photons resulting from up-scattering out of the driven mode.

Although the line narrowing was reported in several other driven cavity systems~\cite{sawant_scirep_2017,bahuleyan_optcont_2025,hauss_prl_2008}, the narrowing observed here results from incoming flux of excitations that are redistributed by intrinsic nonlinearity. Also our results may be contrasted to the Josephson parametric amplifier setting~\cite{eichler_epj-qt_2014}, where one typically considers a single or a pair of modes, and upon increasing driving observes a parametric gain that can be interpreted as a decreased linewidth. In contrast, our setting has a large number of modes, complex fluxes of excitations between them that are determined by energy mismatch and matrix elements. We also do not reach the point of parametric instability where gain would overturn the losses, leading to negative values of the full (not excess) linewidth.  Observing this mode narrowing is feasible with current experimental capabilities, although requires a certain interval of pumping strength. 

\subsection{Memory of drive configuration in NESS}\label{subsec:strongdrive}
After understanding the qualitative features of the NESS and mode linewidth at weak driving, we proceed with the situation when the system is brought far from equilibrium. To this end, we keep temperature fixed to $T=0.1\,\text{K}$, we enlarge the number of pumped modes, ${\mathbb K} = \{1,2,\dots, 19\}$, do not impose any photon flux for higher energy modes, and consider larger values of pump intensity up to $\alpha=80$. Converging the kinetic equation towards the steady state is more computationally demanding for large $\alpha$, as it requires progressively smaller time-steps. 

\begin{figure}[t] 
\centering
        \includegraphics[width=\linewidth]{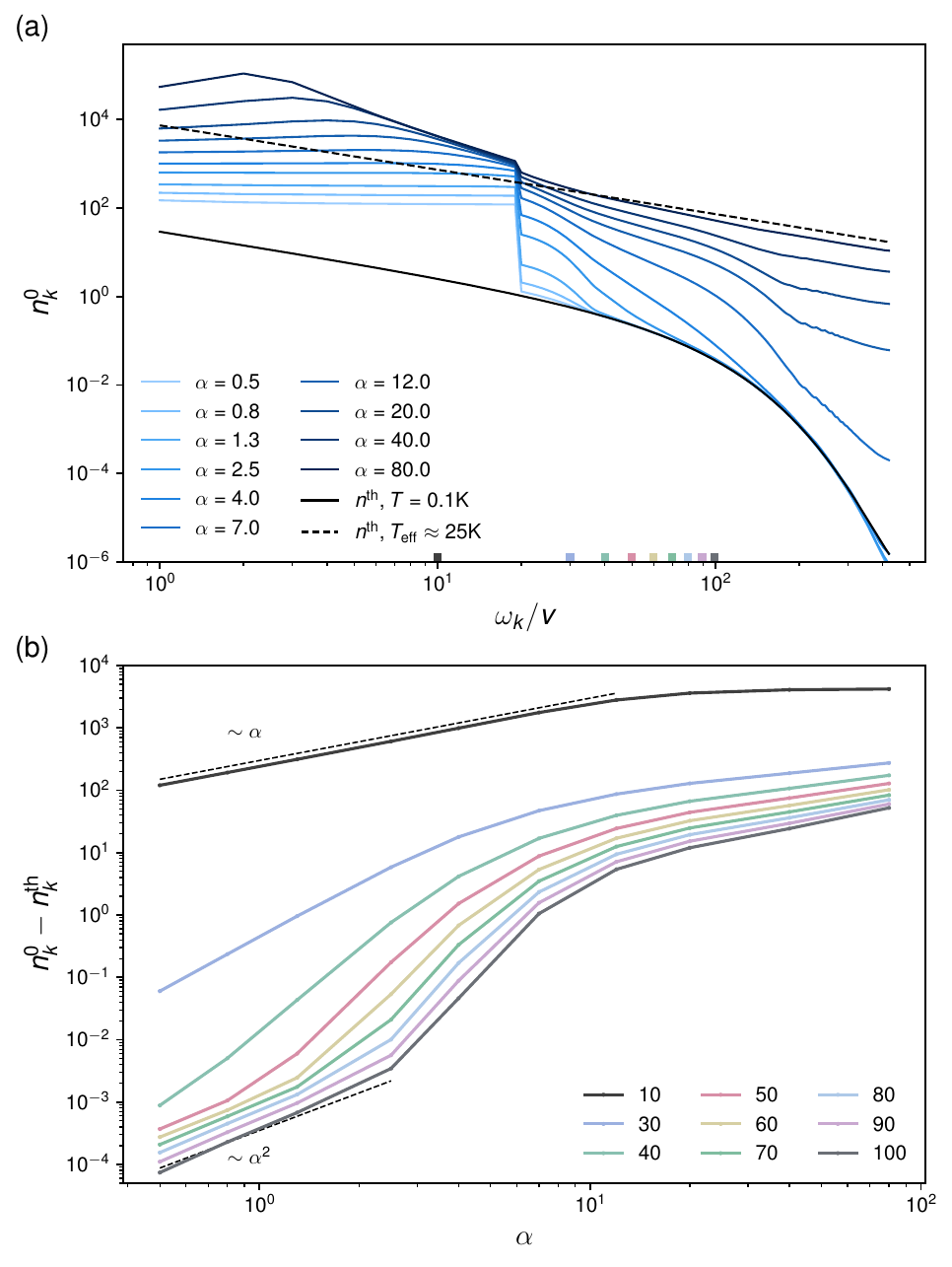}
        \caption{(a) NESS distribution $n^{0}_k$ resulting from pumping modes 1-19 at different intensities, as specified in the legend. On the $x$-axis we show $\omega_k/v$, which approximately equals to $k$ for small mode numbers, however takes into account the spectrum nonlinearity at larger mode numbers. Due to the high effective temperature, the Bose distribution on this scale is a straight line, $n_k\propto 1/\omega_k$. Colored dots mark the modes considered in (b). (b) Difference between the NESS distribution $n_k^0$ and the thermal occupation $n_k^\text{th}$ as a function of the pump intensity $\alpha$. Line colors correspond to the modes identified in panel (a) and specified in the legend.}\label{fig:kin1}
\end{figure}

Figure \ref{fig:kin1}(a) shows the evolution of the steady state distribution function upon changing the intensity $\alpha$. At small $\alpha$, similarly to the previous pumping configuration, the distribution function deviates from its equilibrium value only for the pumped modes and in their vicinity. For larger $\alpha \geq 2.5$, $n^0_k$ becomes substantially different already for the first ~100 modes. For even larger $\alpha\geq 7$, we see that the distribution function deviates strongly for all modes, and additionally gets an upward curvature, in contrast to the Bose-Einstein distribution $n^\text{th}_k$ -- that after an initial power-law part has an approximately exponential decay with $k$, provided we ignore the dispersion's curvature.

The qualitative change in behavior of the steady state distribution function $n^0_k$ with $\alpha$ may be attributed to an enhancement of intrinsic scattering, solicited by the increase in occupation number of modes due to pumping. Na\"ively, the regime of dominant intrinsic scattering -- where the parameter $\tau$ introduced in Section~\ref{subsec:parameters} becomes large -- may be expected to lead to an emergent thermal behavior: excitations entering the chain experience numerous intrinsic scattering events, and eventually reach an effective thermal equilibrium with a temperature set by the balance of incoming and radiated energy fluxes.

The data shown for $n_k^0$ in Fig.~\ref{fig:kin1}(a) at the largest values of $\alpha$ show a power-law-like dependence of $n^0_k$ on $k$ extending over a relative broad range of mode numbers $k$ in the interval $[50,300]$. The dependence is approximately consistent with a simple power-law $n^0_k \propto 1/k$, that can be interpreted as a thermal equilibrium with elevated temperature (expanding the Bose distribution gives $n^0_k \propto 1/\omega_k\approx 1/k$). However, $n^0_k$ eventually develops a convex segment, that is inconsistent with such interpretation, as a Bose function is expected to have opposite curvature. This can be further checked by plotting the Bose distribution at the effective temperature corresponding to the same total energy in Eq.~\eqref{eqn:conservation_laws}, which is found to be $T_\text{eff}\approx 25\,\text{K}$. Such value reflects the significant amount of energy introduced in the system through pumping. 

We notice that the results shown in Fig.~\ref{fig:kin1} are obtained by simple solution of the kinetic equation without self-consistent determination of linewidths. However, in the present case, when the excess linewidth exceeds the original broadening determined by terminal coupling and environment, keeping linewidth fixed is not justified anymore. Applying the same phenomenological linewidth update as in Ref.~\cite{bubis_sciadv_2026}, we obtain qualitatively different results for the NESS distribution function $n^0_k$ (not shown), with a more rapid power-law decay with $k$, where the exponent ranges between $2$ and $4$.

Such approximate power-law dependence of $n^0_k$  on $k$ may be suggestive of the realization of a non-thermal fixed point in the regime of strong driving~\cite{berges_prl_2008}. The non-thermal fixed point corresponds to self-similar solutions of kinetic equations in the driven regime, that were also reported in the context of weak turbulence governing the behavior of shallow water waves, described by a similar kinetic equation~\cite{nazarenko_book_2011}. We leave a more detailed investigation of the strongly driven regime and the emergent power-law behavior of $n_k^0$ with $k$ to future work, and meanwhile focus on a better understanding of the crossover to such different steady state. 

\begin{figure}[t] 
\centering
        \includegraphics[width=\linewidth]{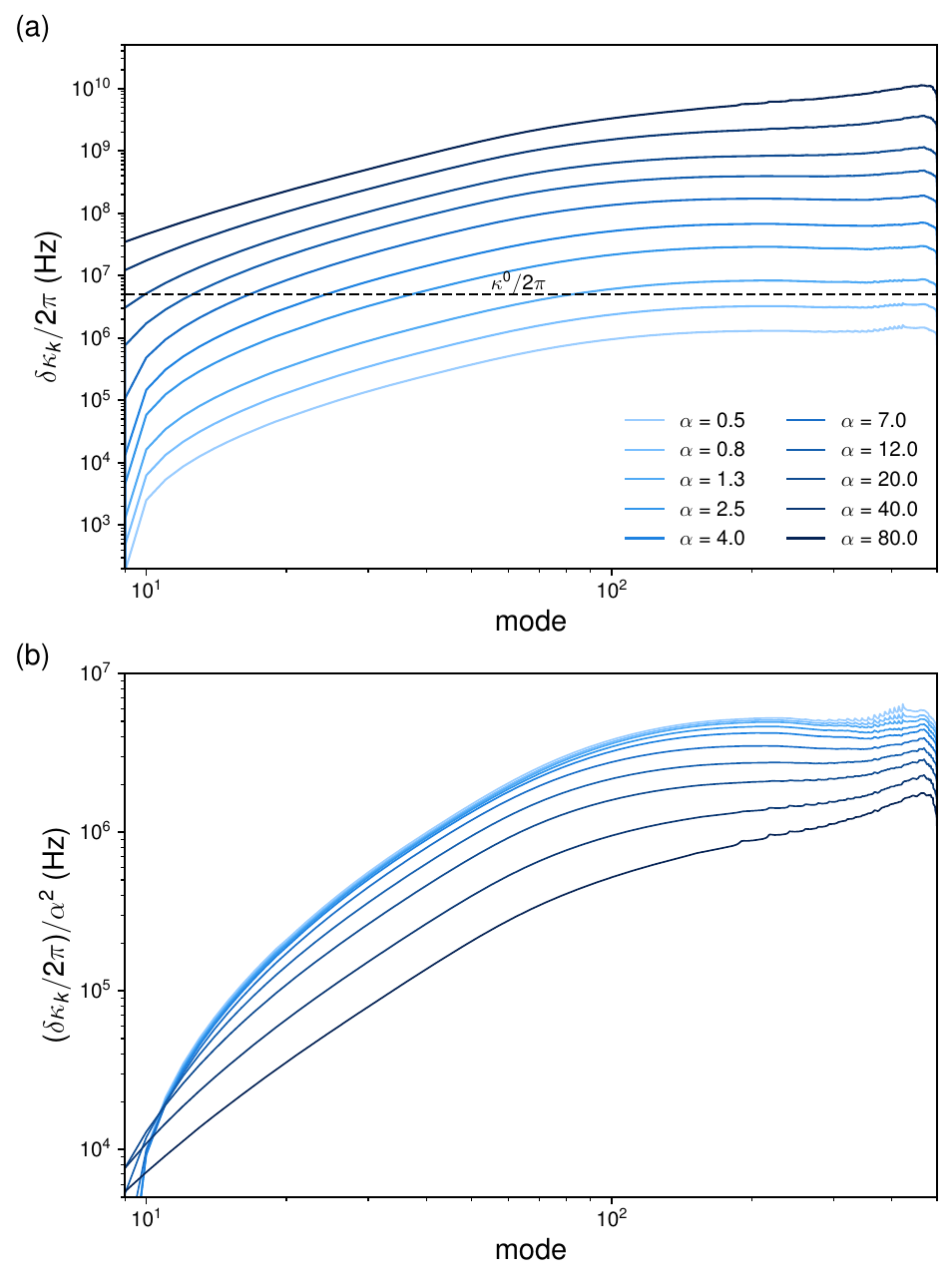}
        \caption{(a) Excess linewidth $\delta\kappa_k$ as a function of mode numbers calculated at NESS in panel (a). The dashed black line indicates the value of bare linewidth $\kappa^0$. Modes with $k<10$ depending on value of $\alpha$ may feature $\delta\kappa_k<0$ due to a prevalence of in-scattering, as discussed in the previous Section, and are therefore not shown. (b) When $\delta\kappa_k$ is below or comparable to the bare linewidth, it scales quadratically with the pump intensity $\alpha$, as shown by the collapse of the three lower pump intensity curves. Colors are the same as in (a).}\label{fig:kin2}
\end{figure}

To understand the crossover to the regime where NESS looses memory of the pumping configuration, we consider the change in occupation of individual non-pumped modes with fixed $k$, $n_k^0$, versus intensity $\alpha$, shown in Fig.~\ref{fig:kin1}(b). For modes far enough from pumped modes, the change in $n^0_k-n^\text{th}_k$ shows a collapse as a function of $\alpha$ to the approximate power-law behavior, $n^0_k-n^\text{th}_k \propto \alpha^w$, with $w\approx 2$. At intermediate values of $\alpha \approx 5-10$ the collapse breaks down and the dependence on $\alpha$ becomes much steeper. Finally, at even larger pumping intensities, we see again an emergent collapse, now with exponent close to 1. However, for modes closer to the pumped set we observe a stronger dependence on $\alpha$. Heuristically, the complex dependence of $n^0_k-n^\text{th}_k$ on $\alpha$ is not surprising, and it can be understood via the minimal number of pump-involving scattering events required to connect a considered mode $k$ to pumped modes. Within a na\"ive description, each such event contributes one power of the population of pumped modes, $n_k\propto \alpha$ for $k\in \mathbb{K}$, so modes reached only through longer scattering cascades exhibit higher-order power laws. Consequently, the scaling exponent does not correlate simply with the energy distance from the pumped mode: a distant mode may be populated by a direct process involving two pumped modes and scale with $w\approx 2$, while a closer mode may require one or more intermediate modes and therefore scale with a higher exponent.

After considering the NESS distribution function, we turn our attention to mode broadening, that is potentially easier to access within experiments. Figure~\ref{fig:kin2}(a) shows the contribution to broadening coming from intrinsic scattering for different values of $\alpha$. A criterion for the  NESS to keep the memory about pumping configuration is
\begin{equation}
    \delta\kappa_k\ll\kappa^0.
\end{equation}
This can be obtained by approximating the NESS as $ n_k^*= (\kappa_{{\rm ex},k}/\kappa^0_k)n^{\rm flux}_k+n_k^{\rm th}$ and expanding the collision integral to obtain the excess linewidth. Intuitively, this criterion implies that the collision timescale $\delta\kappa_k^{-1}$ is much longer than the damping timescale $(\kappa_k^0)^{-1}$, the system retains memory of the drive configuration. When this criterion is violated, the intrinsic scattering increases, and the system may potentially enter the (non-)thermal fixed point regime. This criterion can be translated into a power-law dependence of $\delta\kappa_k$ on the drive strength $\alpha$. As shown in Sec.~\ref{subsec:vnadrive}, at low drive intensity the relevant processes involve two or three pumped modes. According to Eq.~\eqref{eqn:currentdecays}, both cases lead to a $\delta\kappa_k \propto  \alpha^2$ scaling. This behavior is clearly visible in Fig.~\ref{fig:kin2}(b), where $\delta\kappa_k$ curves at the four lowest intensities collapse onto a single curve when rescaled by $\alpha^2$. At higher drive strengths, this scaling breaks down and the dependence becomes approximately linear in $\alpha$, as consistent with the redistribution of energy among the entire set of modes.

To summarize, in the strong driving regime we predict a crossover in the behavior of the linewidth and occupation numbers as a function of pumping intensity. At the highest intensities, we expect the onset of a qualitatively different NESS distribution, in which the memory of the pumping configuration is completely washed out. Determining whether this distribution takes an approximately thermal form or instead corresponds to a non-thermal fixed point remains an interesting avenue for future work.

\section{Summary and discussion}\label{sec:disc}
In this paper, we studied the internal relaxation of plasma modes in a JJ chain through two-into-two scattering arising from multi-mode nonlinearities. These intrinsic relaxation processes lead to a measurable excess linewidth both in the presence and in the absence of an external drive. Despite arising at the same order as Kerr nonlinearities~\cite{weissl_prb_2015,krupko_prb_2018}, two-into-two scattering, also knows as four-wave mixing, remained relatively unexplored until recently~\cite{bubis_sciadv_2026}. Yet, the problem is interesting for several reasons. From a theoretical point of view, our starting Hamiltonian is referable to a Luttinger liquid as a limiting case, where understanding effects of inelastic decay appears as a long-standing problem~\cite{lin_prl_2013,imambekov_rmp_2012}. More practically, including two-into-two scattering leads to the implementation of a Bose-Hubbard model with interactions that are non-local in momentum space. This corresponds to a general description of nonlinear multi-mode cavities, potentially relevant for quantum optics and suitable to different types of resonators. In this framework, the coherence of JJ chains and their response to external driving is a specific question relevant for the broad range of implementations of multi-mode circuits quantum electrodynamics.

In the first part of this work, we discussed the impact of a finite linewidth and the discreteness of plasmon modes on their lifetime at thermal equilibrium. We first connected to previous literature which assumed a continuum of zero-linewidth modes~\cite{lin_prl_2013, bard_prb_2018}, and discussed in which limit their results are recovered. These works considered the only possible on-shell process, which involves a large momentum difference between initial and final states and leads to an excess linewidth $\delta\kappa_k\sim T k^4$. In our work, we showed that such process is hard to observe once modes are discrete and broadened, as a stringent set of conditions is needed. Due to broadening, deviations from such resonance are allowed, changing the expected scaling to $\delta\kappa_k\sim T^2 k^4$. However, the resulting decay rate is predicted to be smaller than bare linewidth and therefore still challenging to observe experimentally. More importantly, broadening allows for another class of processes which would otherwise be off-resonant, with a small momentum difference between initial and final states. This features a power law $\delta\kappa_k\sim T^3 k^2$, and can be comparable to bare linewidth at high temperature and mode number. In order to give an intuitive explanation of all scalings, we provided closed-form analytical expressions which match scalings predicted by numerical results.

In the second part we investigated the out-of-equilibrium regime and NESS by numerically solving the kinetic equation for modes' occupation numbers.  We considered several pumping configurations that involve the incoherent pump of the low-lying modes, along with pumping of an isolated high-lying mode, as well as configurations where only high- or low-lying modes are pumped. Connecting to our previous analysis of different processes, we studied the system's equilibration in presence of pumping of low-frequency modes and a higher frequency one. Keeping track of all collisions where two modes out of four are in the set of pumped states, we found that pumped photons easily scatter to nearby modes, which correspond to processes with a small momentum transfer discussed above. Interestingly, at low pump intensities, we found that the on-shell processes which were hardly visible at thermal equilibrium appear as well-defined peaks of enhanced population in the NESS distribution function. These peaks are eventually washed away at sufficiently high pump power. Persisting peaks are observed instead at frequencies multiple of the single pumped mode. This phenomenology seems reminiscent of cascaded decays observed in~\cite{bubis_sciadv_2026}.

An additional notable phenomenon occurring in the NESS is mode narrowing due to the incoming flux of excitations. We found that pumping a single high-energy mode is sufficient to observe such phenomenon, and estimated the pump intensity required for mode narrowing. Finally, going deeper into the non-equilibrium regime, we analyzed the evolution of NESS distribution function with increasing pump intensity for incoherent low-frequency pump.  We found that the NESS distribution function looses memory of the pumping configuration when excess linewidth from intrinsic scattering exceeds the bare linewidth, $\delta\kappa_k>\kappa^0$. Upon increasing the pumping power, the decay rates of individual modes initially increases quadratically with the pump intensity, $\delta\kappa_k\sim\alpha^2$. A deviation from this dependence marks the crossover between the regimes dominated by extrinsic and intrinsic relaxation, respectively, and should be observable in experiments.

Our work also leaves a number of open questions, inviting future theoretical studies. In particular, we used the kinetic equation framework, ignoring coherences between different modes and considering mode occupations as the only parameters. This approximation may break down for strong drives, where the strong coupling between modes may require a self-consistent treatment of their linewidth and other parameters. A phenomenological treatment of self-consistent linewidths was introduced in Ref.~\cite{bubis_sciadv_2026}, however a more rigorous derivation of such approach remains an interesting avenue for future work. 

Another example where the simplified kinetic equation may be not capturing all physical processes are cases of coherent driving of several modes, that was also demonstrated experimentally~\cite{bubis_sciadv_2026}. Such coherent driving in presence of two-into-two scattering may lead to the creation of non-trivial quantum coherences and squeezed states, that could be explored theoretically and experimentally. Also, the role of scattering processes involving phase slips and their potential activation in NESS remains to be understood. 

Finally, the microwave energy scale of excitations in JJ chains  potentially allows access to not only to the NESS, but also to \emph{transient} phenomena. This motivates the theoretical study of occupation dynamics and other observables in response to a sudden or gradual onset of driving. From a broader perspective, the experimental and theoretical characterization of JJ chains using the available control space for pumping and creation of NESS and dynamics, may allow to establish them as a platform for quantum simulation of driven-dissipative Bose-Hubbard model~\cite{houck_natphys_2012}.

\begin{acknowledgments}
We acknowledge useful discussions with Anton Bubis, Vladimir Manucharyan, Alexander Mirlin and Thomas Scaffidi. We are grateful to Eli Levenson-Falk and Alexander Mirlin for their feedback on the manuscript. This research was supported in part by grant NSF PHY-2309135 to the Kavli Institute for Theoretical Physics (KITP), and by the Austrian Science Fund (FWF) SFB F86.
\end{acknowledgments}

\appendix

\section{Derivation of matrix element}\label{appendixA}
Here, we derive explicitly the four-wave mixing matrix element in Eq.~\eqref{eqn:boxmatrixel}. We start from the fourth order expansion term in Eq.~\eqref{eqn:quartic},
\begin{equation}
  H^{(4)}=-\frac{E_J}{4!}\sum_{n=0}^{N-1}(\theta_{n+1}-\theta_n)^4,
\end{equation}
where we extended the sum to run from $0$ to $N-1$ since $\theta_0=\theta_N=0$. Substituting Eq.~\eqref{eqn:thetaLutt}, we get
\begin{widetext}
\begin{align}
  H^{(4)}=&-\frac{E_J}{24K_g^2\pi^2}\sum_{n=0}^{N-1}\left\{\sum_{k=1}^{N-1}\frac{1}{\sqrt{k}}\left[
  \sin{\left(\frac{\pi k(n+1)}{N}\right)}-\sin{\left(\frac{\pi kn}{N}\right)}
  \right](\hat{a}_k+\hat{a}^\dagger_k)
  \right\}^4\notag\\
  =&-\frac{E_J}{24K_g^2\pi^2}\sum_{n=0}^{N-1}\left\{\sum_{k=1}^{N-1}\frac{2}{\sqrt{k}}
  \sin{\left(\frac{\pi k}{2N}\right)}\cos{\left[\frac{\pi k}{N}\left(n+\frac{1}{2}\right)\right]}
  (\hat{a}_k+\hat{a}^\dagger_k)\right\}^4\notag\\
  =&-\frac{4E_g}{3\pi^2}\sum_{k,p,q_1,q_2=1}^{N-1}\frac{1}{\sqrt{kpq_1q_2}}\sin{\left(\frac{\pi k}{2N}\right)}\sin{\left(\frac{\pi p}{2N}\right)}\sin{\left(\frac{\pi q_1}{2N}\right)}\sin{\left(\frac{\pi q_2}{2N}\right)}(\hat{a}_k+\hat{a}^\dagger_k)(\hat{a}_p+\hat{a}^\dagger_p)(\hat{a}_{q_1}+\hat{a}^\dagger_{q_1})(\hat{a}_{q_2}+\hat{a}^\dagger_{q_2})\notag\\
  &\sum_{n=0}^{N-1}\cos{\left[\frac{\pi k}{N}\left(n+\frac{1}{2}\right)\right]}\cos{\left[\frac{\pi p}{N}\left(n+\frac{1}{2}\right)\right]}\cos{\left[\frac{\pi q_1}{N}\left(n+\frac{1}{2}\right)\right]}\cos{\left[\frac{\pi q_2}{N}\left(n+\frac{1}{2}\right)\right]}\notag\\
  =&-\frac{E_g\pi^2}{96N^3}\sum_{k,p,q_1,q_2>0}\sqrt{kpq_1q_2}(\hat{a}_{k}^{\dagger} + \hat{a}_{k})(\hat{a}_{p}^{\dagger} + \hat{a}_{p})(\hat{a}_{q_1}^{\dagger} + \hat{a}_{q_1})(\hat{a}_{q_2}^{\dagger} + \hat{a}_{q_2})\sum_{s_1,s_2,s_3=\pm}\delta_{k+s_1p+s_2q_1+s_3q_2,0},
\end{align}
which is Eq.~\eqref{eqn:nl_2ndq}. In the last row we expanded sine functions assuming $k\ll N$, and we used
\begin{equation}
    \sum_{n=0}^{N-1}\prod_{i=1}^4\cos{\left[\frac{\pi k_i}{N}\left(n+\frac{1}{2}\right)\right]}
    =\frac{N}{8}\sum_{s_1,s_2,s_3=\pm}\delta_{k_1+s_1k_2+s_2k_3+s_3k_4,0}.
\end{equation}
The four-wave mixing matrix element is given by
\begin{multline}
    \mathcal{K}_{k,p,q_1,q_2}=\langle 0| \hat{a}_k \hat{a}_pH^{(4)} \hat{a}_{q_1}^\dagger \hat{a}_{q_2}^\dagger |0\rangle\\
    =-\frac{E_g\pi^2}{96N^3}\sum_{i,j,l,r>0}\sqrt{ijlr}\langle 0|\hat{a}_k \hat{a}_p(\hat{a}_{i}^{\dagger} + \hat{a}_{i})(\hat{a}_{j}^{\dagger} + \hat{a}_{j})(\hat{a}_{l}^{\dagger} + \hat{a}_{l})(\hat{a}_{r}^{\dagger} + \hat{a}_{r}) \hat{a}_{q_1}^\dagger \hat{a}_{q_2}^\dagger|0\rangle\sum_{s_1,s_2,s_3=\pm}\delta_{i+s_1j+s_2l+s_3r,0}\\
    =-\frac{E_g\pi^2}{4N^3} \sqrt{kpq_1q_2}\sum_{s_1,s_2,s_3=\pm}\delta_{k+s_1p+s_2q_1+s_3q_2,0},
\end{multline}
which is exactly Eq. \eqref{eqn:boxmatrixel}.

\section{Kerr coefficients}\label{appendixB}
In this Appendix, we derive Eq. \eqref{eqn:kerrcoeff}. In order to obtain Kerr coefficients, we need to isolate terms in $H^{(4)}$ where mode indices are all equal or equal in pairs. $H^{(4)}$ with indices equal in pairs, responsible for cross-Kerr effect, is easily found to be
\begin{equation}
     H^{(4),\text{ cross}} =-\frac{\hbar}{2}\sum_{k,p>0,p\neq k}\mathcal{K}^{k\neq p}_{k,p}\hat{n}_{k}\hat{n}_{p}
     -\frac{\hbar}{2}\sum_{k,p>0,p\neq k}\mathcal{K}^{k\neq p}_{k,p}\hat{n}_{k}-\frac{\hbar}{8}\sum_{k,p>0,p\neq k}\mathcal{K}^{k\neq p}_{k,p},
\end{equation}
where $\hat{n}_k=\hat{a}^\dagger_k\hat{a}_k$ and
\begin{align}
    \mathcal{K}^{k\neq p}_{k,p}=\frac{\pi^2 E_g}{2N^3\hbar}kp=\frac{\pi^2 E_g}{2N^3\hbar}\frac{\omega_k\omega_p}{v^2}=\frac{\hbar}{4NE_J}\omega_k\omega_p.
\end{align}
If all mode indices are equal, we get
\begin{equation}
     H^{(4),\text{ self}} =-\frac{3\hbar^2}{32E_JN}\sum_{k>0}\omega_k^2 \hat{n}_k\hat{n}_k
     -\frac{3\hbar^2}{32E_JN}\sum_{k>0}\omega_k^2 \hat{n}_k-\frac{3\hbar^2}{64E_JN}\sum_{k>0}\omega_k^2.
\end{equation}
Collecting all terms,
\begin{multline}
H^{(2)}+H^{(4)}=\sum_{k>0}\hbar\omega_k\left(\hat{n}_k+\frac{1}{2}\right)-\frac{\hbar}{2}\sum_{k>0}\sum_{p>0,p\neq k}\mathcal{K}^{k\neq p}_{k,p}\hat{n}_{k}-\frac{3\hbar^2}{32E_JN}\sum_{k>0}\omega_k^2 \hat{n}_k\\
-\frac{\hbar}{2}\sum_{k>0}\sum_{p>0,p\neq k}\mathcal{K}^{k\neq p}_{k,p}\hat{n}_{k}\hat{n}_{p}-\frac{3\hbar^2}{32E_JN}\sum_{k>0}\omega_k^2 \hat{n}_k\hat{n}_k\\
-\frac{\hbar}{8}\sum_{k>0}\sum_{p>0,k\neq p}\mathcal{K}^{k\neq p}_{k,p}-\frac{3\hbar^2}{64E_JN}\sum_{k>0}\omega_k^2.
\end{multline}
\end{widetext}
In the expression above, terms in the first row give a renormalization of the bare mode frequency, $\omega_k\rightarrow \omega'_k$, in the zero occupation number limit. Terms in the second row give cross-Kerr and self-Kerr corrections for non-zero occupation number. We can further rewrite the latter as
\begin{align}
    &-\frac{\hbar}{2}\sum_{k>0}\sum_{p>0,p\neq k}\mathcal{K}^{k\neq p}_{k,p}\hat{n}_{k}\hat{n}_{p}-\frac{3\hbar^2}{32E_JN}\sum_{k>0}\omega_k^2 \hat{n}_k\hat{n}_k\notag\\
    =&-\frac{\hbar}{2}\left[\sum_{k>0}\sum_{p>0}\underbrace{\left(\frac{1}{2}-\frac{\delta_{k,p}}{8}\right)\frac{\hbar\omega_k\omega_p}{2NE_J}}_{\equiv\mathcal{K}_{k,p}}\hat{n}_{k}\hat{n}_{p}\right],\label{eqn:kerrcoeffbox}
\end{align}
obtaining expression in the Eq. \eqref{eqn:kerrcoeff} in the main text.

\section{One-to-three scattering}\label{appendixC}
Here we verify that $1\to 3$ scattering arising from terms $\hat{a}_k\hat{a}_p\hat{a}_{q_1}\hat{a}_{q_2}^{\dagger}$ is energetically much less favorable than $2\to 2$ scattering that conserves the number of bosons. As discussed in the main text, for $2\to 2$ scattering there are well-known on-shell configurations satisfying energy-momentum conservation without mode broadening (see \cite{lin_prl_2013, bard_prb_2018}), as well as favorable near-resonant cases with small momentum transfer. In contrast, for $1\to 3$ scattering we did not find comparable relevant processes, as we show below in further details. This is also mentioned in \cite{houzet_prl_2019} and \cite{lin_prl_2013}. Thus, although $1\to 3$ terms arise at the same nonlinear order as $2\to 2$, their reduced phase space and weaker resonance justify neglecting them in our analysis.

To justify the smallness of $1\to 3$ scattering, we consider the expanded dispersion relation in Eq. \eqref{eqn:mirlindisp}, $\omega_k=vk(1-\xi k^2)$, we adopt our convention with both positive and negative modes, and assume the probed mode to have positive sign, $k>0$. For the $2\to 2$ case, the single scattering configuration where momentum and energy are conserved, even in the absence of broadening of the modes, is $k,p\to q_1,q_2$, with $q_1,q_2>0$ and $p=-\frac{3}{2}\xi kq_1q_2<0$ -- process \textit{(i)}. Besides this, the other favorable configuration is process \textit{(ii)}, where $q_1\approx k, q_2\approx p$ and $p>0$ (the process where $k=q_1$, $p=q_2$ has to be excluded). We want to check now what are the energetically favorable configurations in the $1\to 3$ case – say $k,p\to q_1,q_2$. Using momentum and energy conservation we get
\begin{align*}
    &k=p+q_1+q_2,\\
    &k(1-k^2)-|p|(1-p^2)-|q_1|(1-q_1^2)-|q_2|(1-q_2^2)=0,
\end{align*}
with $p,q_1,q_2\gtrless 0$. Among the possible sign combinations, $p,q_1,q_2>0$ or $p,q_1<0,\,q_2>0$ or $p<0,\,q_1,q_2>0$, only the last one admits exact solution in principle, but it is subjected to a condition over $\xi$ and $k$, which is not fulfilled within our range of mode numbers and for typical values of $\xi$.

\begin{figure*}[tb]
\centering
        \includegraphics[width=\textwidth]{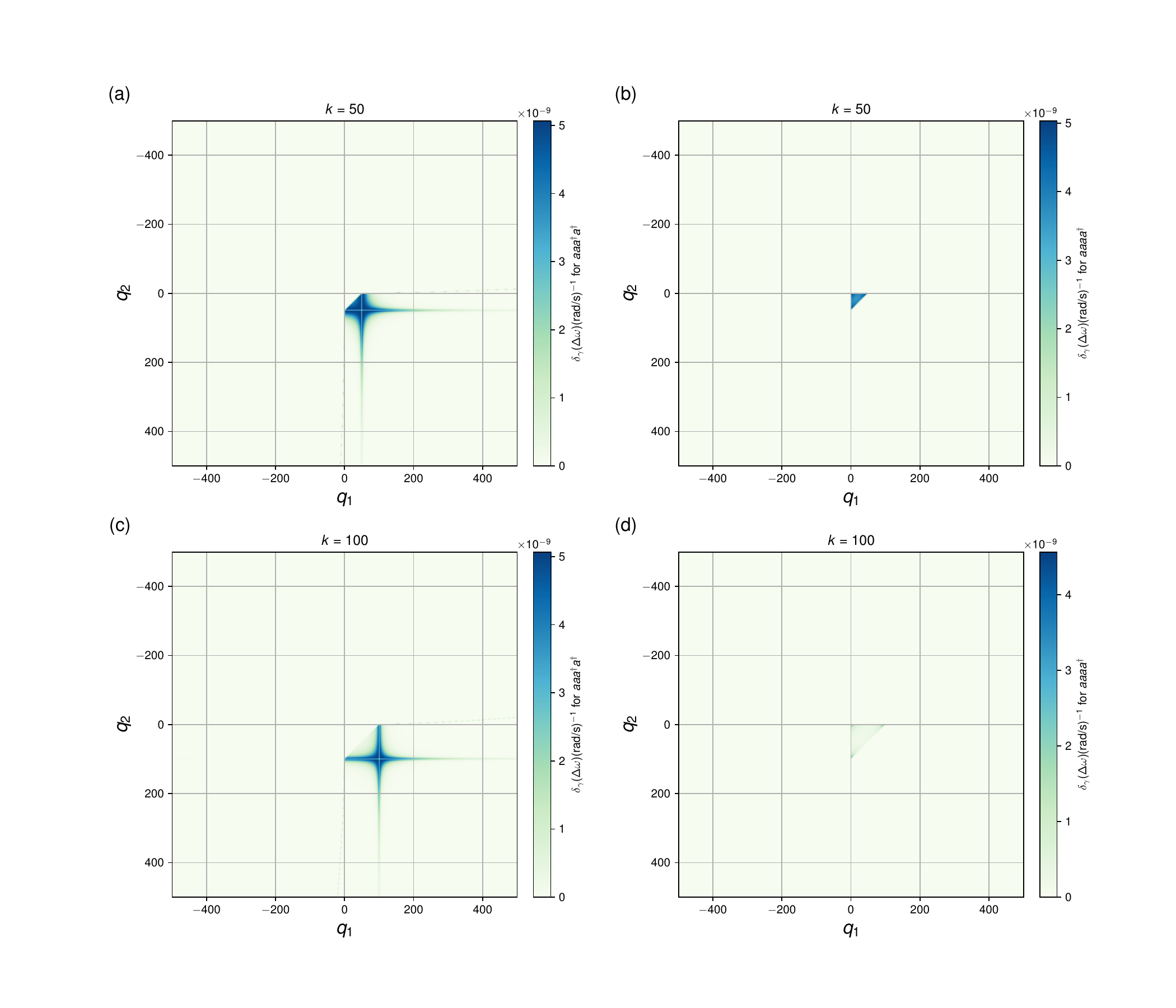}
        \caption{Comparison between $\delta_\gamma(\Delta\omega)$ assuming $2\to2$ scattering (left panels) and $1\to3$ scattering (right panels). We consider mode $k=50$ in panels (a)-(b), and mode $k=100$ in panels (c)-(d). In $1\to3$ scattering, lower values of $\delta_\gamma(\Delta\omega)$ and a reduced number of relevant decay channels are observed, consistently with the analytical discussion. In all panels, $\kappa^0/2\pi=5\,\text{MHz}$.}\label{fig:aaaa+}
\end{figure*}

To further numerically support our claim that $1\to 3$ process is not important, we plot numerically the broadened density of states, $\delta_\gamma(\Delta\omega)$ (Eq. \eqref{eqn:dos}) as a 2d-plot, as a function of $q_1,q_2$, and for a fixed $k$, see Fig. \ref{fig:aaaa+}. For the  $1\to 3$ scattering, we see features only when both $q_1$ and $q_2$ are positive. This could be expected since, if all modes have the same sign, energy conservation is violated at the level of cubic correction $\propto\xi$. Instead, mixed signs introduce an energy violation even at the zeroth order in $\xi$.

Comparing the plots on the left ($2\to 2$ case) to right ($1\to 3$ case), we observe lower values of $\delta_\gamma(\Delta\omega)$ and a reduced number of relevant decay channels, in line with analytical derivation. To summarize, even though processes such as $\hat{a}_k\hat{a}_p\hat{a}_{q_1}\hat{a}_{q_2}^{\dagger}$ arise at the same nonlinear order as $\hat{a}_k\hat{a}_p\hat{a}^\dagger_{q_1}\hat{a}_{q_2}^{\dagger}$, the former is less resonant than the latter, and thus neglected in our analysis.

\section{Sixth order nonlinear term}\label{appendixD}
In order to justify that higher orders in the nonlinear Hamiltonian are negligible in our analysis, below we derive explicitly the sixth order contribution to decay rates, assuming a $3\to3$ scattering process, and verifying that it does not affect our results.

\begin{figure}[t] 
        \includegraphics[width=\columnwidth]{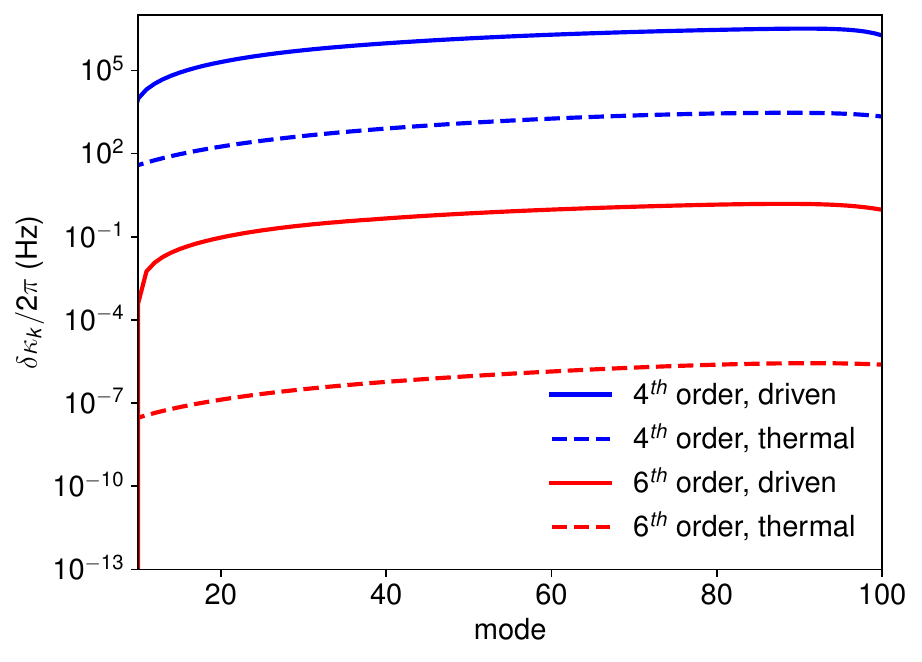}
        \caption{Excess linewidth $\delta\kappa_k$ arising from $2\to2$ scattering (blue) and $3\to3$ scattering (red). Dashed lines correspond to thermal equilibrium, solid lines to the case where first 19 modes are driven with $\alpha=1$.}\label{fig:sixth}
\end{figure}

The sixth order reads as
\begin{equation}
 H^{(6)}=\frac{E_J}{6!}\sum_{n=1}^{N-2}(\theta_{n+1}-\theta_n)^6.
 \end{equation}
Similarly to what we did for the fourth order, we can expand the phase field $\theta_n$ and calculate the matrix element
\begin{widetext}
\begin{multline}
    \mathcal{K}_{k,p_1,p_2,q_1,q_2,q_3}=\langle 0| \hat{a}_{k} \hat{a}_{p_1}\hat{a}_{p_2}H^{(6)} \hat{a}_{q_1}^\dagger \hat{a}_{q_2}^\dagger\hat{a}_{q_3}^\dagger |0\rangle
    =\frac{\pi^3{E_g^{3/2}}}{2^{7/2}N^5 E_J^{1/2}}\sqrt{kp_1p_2q_1q_2q_3}\sum_{s_1,\ldots,s_5=\pm}\delta_{k+s_1p_1+s_2p_2+s_3q_1+s_4q_2+s_5q_3,0}.
\end{multline}
Although the matrix element squared in the sixth order scattering is much smaller than in the fourth order, the phase space is different in the two cases. For a conclusive comparison, we fully calculate the decay rate $\delta\kappa$. We first need to generalize the collision integral. This is found to be
\begin{multline}
I_{k}[n]= \frac{1}{12} \sum_{q_1,q_2,q_3,p_1,p_2} W_{q_1,q_2,q_3 \to k,p_1,p_2} \bigg\{
(1 + n_{p_1})(1 + n_{p_3})n_{q_1} n_{q_2} n_{q_3}  \\
+ n_k \Big[
n_{q_1} n_{q_2} n_{q_3} - n_{p_1} n_{p_2} 
+ n_{q_1} n_{q_2} n_{q_3} (n_{p_1} + n_{p_2}) - n_{p_1} n_{p_2} (n_{q_1} + n_{q_2} + n_{q_3}) 
- n_{p_1} n_{p_2} (n_{q_1} n_{q_2} + n_{q_1} n_{q_3} + n_{q_2} n_{q_3})
\Big]
\bigg\},
\end{multline}
where the prefactor 1/12 accounts for the indistinguishability of the three bosons in one state ($q_1,q_2,q_3$) and of two in the other state ($p_1,p_2$) -- the sixth boson $k$ is not summed -- and
\begin{equation}
     W_{q_1,q_2,q_3\to k,p_1,p_2}=2\pi |\mathcal{K}_{q_1,q_2,q_3,k,p_1,p_2}|^2\delta(\omega_{q_1}+\omega_{q_2}+\omega_{q_3}-\omega_k-\omega_{p_1}-\omega_{p_2}).
\end{equation}
As we did before with fourth order nonlinearities, we exclude self-decays. Finally, the decay rate originating from $3\to3$ collisions, which we denote as $\delta\kappa_k^{(6)}$, is given by
\begin{multline}
    \delta\kappa_k^{(6)} =\frac{1}{12} 2\pi\frac{\pi^6E_\text{g}^3}{2^7N^{10}E_J}\sum_{p_1,p_2,q_1,q_2,q_3}kp_1p_2q_1q_2q_3\left[n_{p_1}n_{p_2}(1+n_{q_1})(1+n_{q_2})(1+n_{q_3})-n_{q_1}n_{q_2}n_{q_3}(1+n_{p_1})(1+n_{p_2})\right]\\
    \delta(\omega_k+\omega_{p_1}+\omega_{p_2}-\omega_{q_1}-\omega_{q_2}-\omega_{q_3})\sum_{s_1,\ldots,s_5=\pm}\delta_{k+s_1p_1+s_2p_2+s_3q_1+s_4q_2+s_5q_3,0}.\label{eqn:deltakappasix}
\end{multline}
\end{widetext}
In Fig. \ref{fig:sixth} we show 
the excess linewidth assuming $n_k$ are the occupation numbers at thermal equilibrium with temperature $0.1$\,K in the absence of driving (dashed lines). For the non-equilibrium steady state, we use the distribution function obtained assuming $\alpha=1.0$ drive of the first 19 modes, and $2\to2$ scattering as the equilibration mechanism (see Fig. \ref{fig:kin1}(a)). However, for a faster convergence, here we considered only the first 100 modes. Using this distribution function, we compare the excess linewidth calculated with Eqs.~\eqref{eqn:currentdecays} and~\eqref{eqn:deltakappasix} for $2\to2$ and $3\to3$ scattering respectively.  We observe that even in the non-equilibrium steady state in presence of driving, the excess linewidth from $3\to3$ scattering is much smaller compared to the one coming from $2\to2$ processes. This justifies our focus on the fourth order nonlinearities in our work.

\section{Quantitative comparison of decay rates with the analytical prediction}\label{appendixF}
In the main text we obtained the scaling dependence of excess linewidth on the mode number and temperature. To obtain these estimates, we introduced a number of approximations used to analytically simplify general expressions. Here we show examples of quantitative comparisons between numerically obtained decay rates and the analytical predictions in Eqs.~\eqref{eqn:mirlintheory}-\eqref{eqn:mirlinofftheory} and~\eqref{eqn:nearbytheory}. This comparison is performed in the parameter regime where the assumptions underlying each expression are valid, even though that regime is not always experimentally accessible. We emphasize that with fewer simplifications, a closed-form expressions for excess linewidth would not have been possible. However, the agreement between the processes isolated in the numerical simulations and the analytical predictions would improve. This confirms our classification of scattering processes, with remaining quantitative discrepancies arising from the approximations made to obtain closed-form expressions.

\subsection{Large momentum transfer, on-shell}
\begin{figure}[b]
        \includegraphics[width=\linewidth]{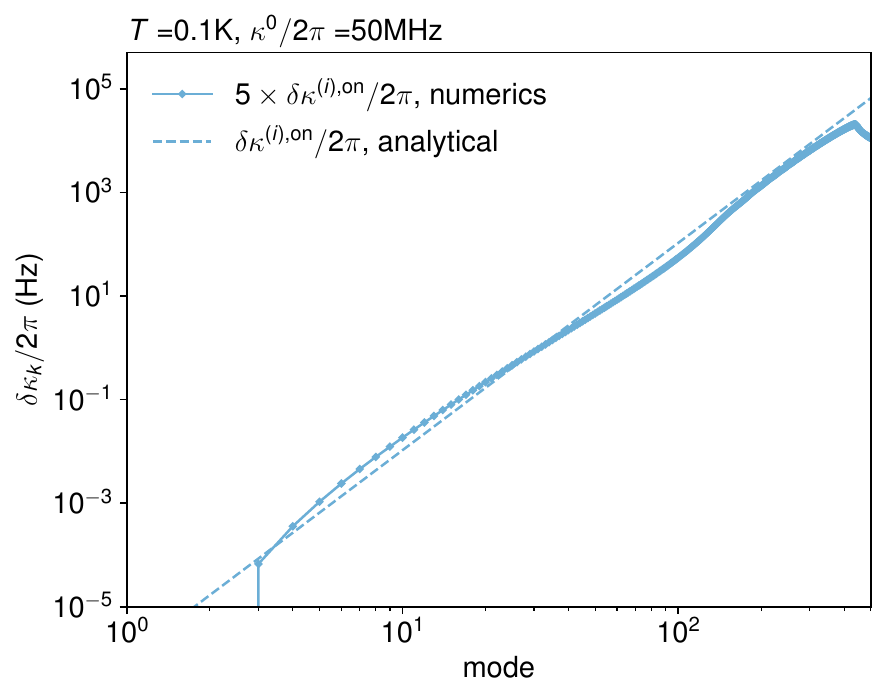}
        \caption{Quantitative comparison with the analytical prediction for processes \textit{(i)} (on-shell contribution), using $T=0.1\,\text{K}$ and $\kappa^0/2\pi=50\,\text{MHz}$ to satisfy the assumptions listed in the main text.}\label{fig:app0}
\end{figure}

In order to satisfy all the assumptions leading to Eq.~\eqref{eqn:mirlintheory}, we choose the following parameters: temperature is set to $T=0.1\,\text{K}$, which satisfies $n_{p^*}\approx T/(v|p^*|)\ge1$ up to mode $k\approx 330$. In addition, we increase a linewidth by a factor of 10, to $\kappa^0/(2\pi)=50\,\text{MHz}$ which now satisfies criteria $2\kappa^0/v>1$ and $\kappa^0\le v|p^*|$ in the whole range $150<k<500$, where the lower bound comes from $|p^*|>1$. The numerical simulation and the analytical prediction shown in Figure~\ref{fig:app0} show similar scaling with the mode number. Quantitatively they differ by a factor $\approx5$.

\begin{figure}[t] 
        \includegraphics[width=\columnwidth]{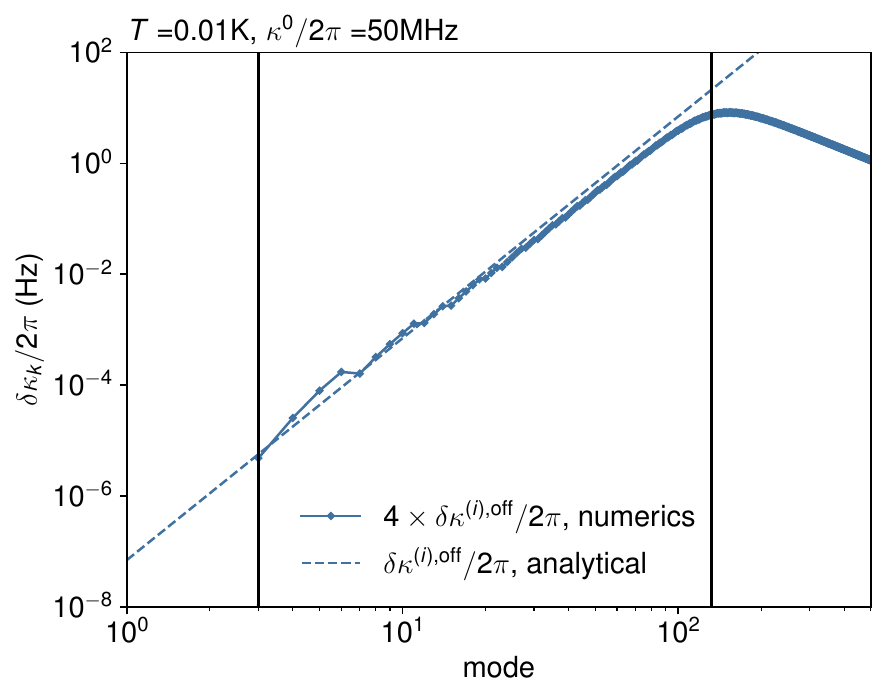}
        \caption{Quantitative comparison with the analytical prediction for processes \textit{(i)} (off-shell contribution), using $T=0.01\,\text{K}$ and $\kappa^0/2\pi=50\,\text{MHz}$ to satisfy the assumptions listed in the main text.}\label{fig:app2}
\end{figure}

\begin{figure}[b] 
        \includegraphics[width=\columnwidth]{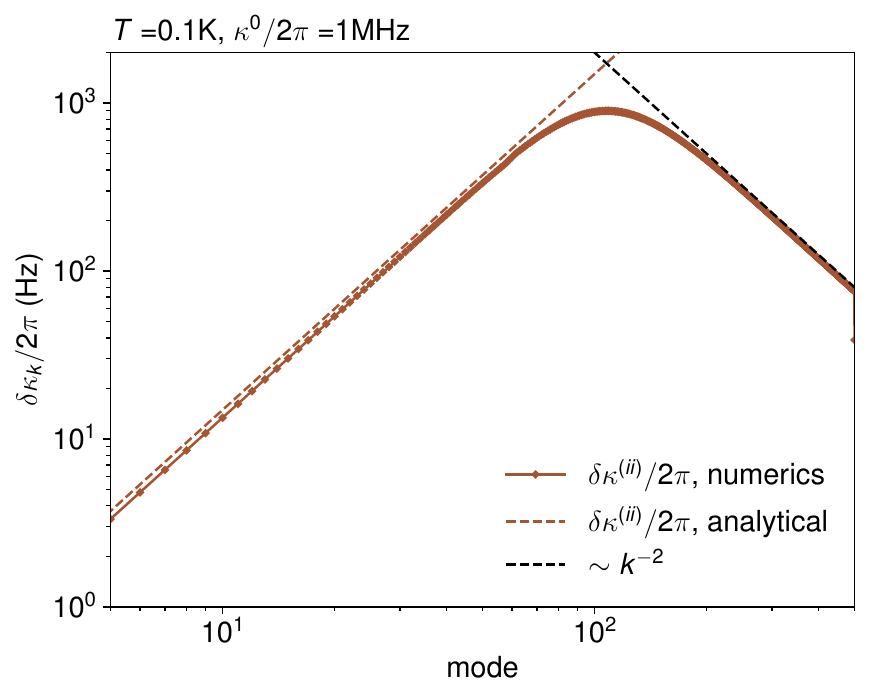}
        \caption{Quantitative comparison with the analytical prediction for processes \textit{(ii)}, using $T=0.1\,\text{K}$, $\kappa^0/2\pi=1\,\text{MHz}$ and $\delta_\text{max}=1$.}\label{fig:(ii)quant2}
\end{figure}

\subsection{Large momentum transfer, off-shell}
For comparing the excess linewidth from large momentum transfer off-shell processes, we use even lower temperature $T=0.01\,\text{K}$, and cut off sums at $p=T/v$. Our analytical expression is derived in the range ${T}/v \ll k\ll \sqrt[3]{{8\kappa^0}/({3v\xi})}$. For $\kappa^0/(2\pi)=50\,\text{MHz}$, this gives $3\ll k\ll 132$. In Fig.~\ref{fig:app2} we compare numerical simulation with the analytically predicted scaling, with vertical lines mark the region of mode numbers where our analytical estimate is applicable. At the same time, $2\kappa^0/v>1$ is still satisfied. This avoids the under-sampling of $\delta_\gamma(\Delta\omega)$ around $p=0$, which would result in a low density of states for $p>0$. The numerical simulation and the analytical prediction differ by a factor $\approx4$. 

\subsection{Small momentum transfer, off-shell}
Finally, to compare the estimate and numerical simulation for small momentum transfer off-shell processes, in Fig.~\ref{fig:(ii)quant2}, we use $T=0.1\,\text{K}$ and even smaller  $\kappa^0/(2\pi)=1\,\text{MHz}$, and we consider only decay to the nearest modes, $\delta_\text{max}=1$ in Eq.~\eqref{eqn:nearbytheory}. For sufficiently large $k$, the approximation for $\delta_{\gamma}(\Delta\omega)$ presented in the main text does not hold, and for sufficiently large $k$, the excess linewidth is expected to scale as $\sim k^{-2}$, that is indeed visible at larger mode numbers. We note that the agreement between the analytical estimate and numerical simulation becomes worse if a $k$-dependent $\delta_{\text{max}}$ is used. The reason is that we neglected the $p$-dependence of the density of states when approximating it by a box of a width $2\delta_{\text{max},k}$.


%

\end{document}